\documentclass[runningheads]{llncs}
\pdfoutput=1
\usepackage{iftex}
\ifpdftex
  \usepackage[T1]{fontenc}
\fi
\usepackage{graphicx}
\usepackage{xcolor}
\usepackage{xurl} 
\usepackage[colorlinks,urlcolor=blue,citecolor=blue]{hyperref}

\usepackage{amsmath,amssymb}
\usepackage{mathtools}
\usepackage{ltl}
\usepackage{listings}
\usepackage{protobuf/lang}  
\usepackage{protobuf/style} 
\usepackage{booktabs}
\usepackage{multirow}
\usepackage{orcidlink}

\newcommand{\tool}{\texttt{NeuroSynt}}
\newcommand{\LTLAP}{\mathit{AP}}

\makeatletter
\newcommand{\printfnsymbol}[1]{%
  \textsuperscript{\@fnsymbol{#1}}%
}
\makeatother

\graphicspath{{graphics/}}

\lstdefinelanguage{json}{
  frame=lines,
  belowcaptionskip=1\baselineskip,
  backgroundcolor=\color{proto_background},
  basicstyle=\color{proto_basic}\scriptsize\ttfamily,
	breaklines=true,
	showstringspaces=false,
	tabsize=2,
  basicstyle=\normalfont\ttfamily,
  numbers=left,
  numberstyle=\scriptsize,
  stepnumber=1,
  numbersep=8pt,
  keywords=[1]{INFO},
  keywordstyle=[1]\color{proto_keyword},
  keywords=[2]{X,G,R,F},
  keywordstyle=[2]\color{proto_type},
  keywords=[3]{inputs,outputs,assumptions,guarantees,semantics},
  keywordstyle=[3]\color{proto_digits},
  literate=
    {:}{{{\color{proto_keyword}{:}}}}{1}
    {,}{{{\color{proto_keyword}{,}}}}{1}
    {->}{{{\color{proto_type}{->}}}}{2}
    {!}{{{\color{proto_type}{!}}}}{1}
    {|}{{{\color{proto_type}{|}}}}{1}
    {\&}{{{\color{proto_type}{\&}}}}{1}
    {\{}{{{\color{proto_keyword}{\{}}}}{1}
    {\}}{{{\color{proto_keyword}{\}}}}}{1}
    {[}{{{\color{proto_keyword}{[}}}}{1}
    {]}{{{\color{proto_keyword}{]}}}}{1},
}

\begin{document}
\title{NeuroSynt: A Neuro-symbolic Portfolio Solver for Reactive Synthesis}
\titlerunning{NeuroSynt: A Neuro-symbolic Portfolio Solver for Reactive Synthesis}
%
\author{Matthias Cosler\inst{1}\orcidlink{0009-0007-3984-0997}
\and
Christopher Hahn\inst{2} \and
Ayham Omar\inst{1}\orcidlink{0009-0009-2711-8993} \and
Frederik Schmitt\inst{1}\orcidlink{0009-0001-7106-3725}
}
\authorrunning{Cosler, Hahn, Omar, Schmitt}
%
\institute{CISPA Helmholtz Center for Information Security, Saarbrücken, Germany\\
\email{\{matthias.cosler, ayham.omar, frederik.schmitt\}@cispa.de}\\
\and
X, the moonshot factory, Mountain View, USA\footnote{Work done while being at Stanford University.}\\
\email{chrishahn@google.com}}
\maketitle              
\begin{abstract}
We introduce \tool, a neuro-symbolic portfolio solver framework for reactive synthesis.
At the core of the solver lies a seamless integration of neural and symbolic approaches to solving the reactive synthesis problem.
To ensure soundness, the neural engine is coupled with model checkers verifying the predictions of the underlying neural models.
The open-source implementation of \texttt{NeuroSynt} provides an integration framework for reactive synthesis in which new neural and state-of-the-art symbolic approaches can be seamlessly integrated.
Extensive experiments demonstrate its efficacy in handling challenging specifications, enhancing the state-of-the-art reactive synthesis solvers, with \tool~contributing novel solves in the current SYNTCOMP benchmarks. 
\end{abstract}
\section{Introduction}

The reactive synthesis problem \cite{churchApplicationRecursiveArithmetic1957} seeks to automatically construct an \emph{implementation} from a system's \emph{specification}.
Rather than delving into the intricate nuances of \emph{how} a system computes, hardware designers can describe \emph{what} the system should achieve and leave implementation details to the synthesis engine. We introduce \tool, a portfolio solver for reactive synthesis that combines the efficiency and scalability of neural approaches with the soundness and completeness of symbolic solvers.

The reactive synthesis problem has seen significant progress in recent years \cite{caludeDecidingParityGames2017,finkbeinerSynthesisHyperproperties2020,finkbeinerBoundedCycleSynthesis2016,pitermanSynthesisReactiveDesigns2006} with active tooling development \cite{abrahamSymbolicLTLReactive2021,cadilhacAcaciaBonsaiModernImplementation2023,ehlersUnbeastSymbolicBounded2011,faymonvilleBoSyExperimentationFramework2017,meyerStrixExplicitReactive2018,khalimovPARTYParameterizedSynthesis2013}, and an annual competition (SYNTCOMP \cite{jacobsReactiveSynthesisCompetition2022}). 
However, applications beyond the competition to an industrial scale are still limited.
The advent of machine learning, empowered by the advancements in deep learning architecture and hardware accelerators, has the potential to drastically increase performance in reactive synthesis.
While deep learning approaches offer efficiency, they lack soundness and completeness guarantees, which are essential to the reactive synthesis problem.

We address this challenge by introducing \tool, a portfolio solver framework for reactive synthesis that aims to bridge the gap between soundness, completeness, and practical efficiency through the combination of state-of-the-art symbolic solver, model-checker, and deep learning techniques.
The integrated neural solver computes candidate implementations while model-checking tools verify the candidate solutions to ensure soundness.
To ensure completeness, the neural solver is backed up by several state-of-the-art symbolic solvers running in parallel.

In particular, our main contribution is the design and open-source implementation of the extensible and efficient portfolio solver.
\tool's design prioritizes extensibility: Its modular architecture facilitates the seamless integration of new models, algorithms, or optimization techniques.
This adaptability ensures that \tool~remains relevant amidst evolving methodologies, providing researchers with a unified platform to experiment, validate, and advance their innovations in the reactive synthesis domain.

Additionally, we contribute an advanced neural solver for reactive synthesis (based on~\cite{schmittNeuralCircuitSynthesis2021}) that handles larger and more complex specifications, improving its performance on real-world instances from SYNTCOMP.

Our results show that deep learning methods can indeed increase the performance of reactive synthesis tools. \tool~provides smaller solutions faster while maintaining soundness and completeness.
Our portfolio solver enhances the performance of the state-of-the-art Strix \cite{meyerStrixExplicitReactive2018} by $31$ samples on the SYNTCOMP 2022 benchmark, and the bounded synthesis tool BoSy \cite{faymonvilleBoSyExperimentationFramework2017} by $152$ samples. Notably, a virtual best solver (VBS) that combines the neural solver with all tools in the SYNTCOMP 2022 competition solves an additional $20$ instances that a VBS without the neural solver could not solve.

\section{Background}

\paragraph{Reactive Synthesis.}
The reactive synthesis problem is a well-known algorithmic challenge, that dates back to Church~\cite{churchApplicationRecursiveArithmetic1957,churchLogicArithmeticAutomata1962} as the problem of automatically constructing an \emph{implementation} from a system's \emph{specification}.
With the decidability findings in 1969 \cite{buchiSolvingSequentialConditions1969} (using games) and 1972 \cite{rabinAutomataInfiniteObjects1972} (using automata), a long history of work on reactive synthesis was initiated. After the introduction of temporal logics in 1977 \cite{pnueliTemporalLogicPrograms1977}, the complexity for LTL reactive synthesis was found to be 2-EXPTIME complete \cite{pnueliSynthesisReactiveModule1989a} but undecidable for distributed systems \cite{pnueliDistributedReactiveSystems1990}.
Since then, many different approaches have been developed (e.g.,~\cite{caludeDecidingParityGames2017,finkbeinerSynthesisHyperproperties2020,finkbeinerBoundedCycleSynthesis2016,pitermanSynthesisReactiveDesigns2006}) and implemented in tools (e.g. \cite{abrahamSymbolicLTLReactive2021,cadilhacAcaciaBonsaiModernImplementation2023,ehlersUnbeastSymbolicBounded2011,faymonvilleBoSyExperimentationFramework2017,khalimovGamebasedBoundedSynthesis,khalimovPARTYParameterizedSynthesis2013,meyerStrixExplicitReactive2018,renkinImprovementsLtlsynt2022}). Moreover, an annual competition, the Reactive Synthesis Competition (SYNTCOMP \cite{jacobsReactiveSynthesisCompetition2022}), associated with the International Conference on Computer Aided Verification (CAV) is organized to track the improvement of algorithms and tooling.

\paragraph*{Linear-time Temporal Logic (LTL).}
LTL extends propositional logic by introducing temporal operators $\LTLuntil$ (until) and $\LTLnext$ (next).
Several additional operators can be derived: $\LTLdiamond \varphi \equiv \mathit{true} \LTLuntil \varphi$ and $\LTLsquare \varphi \equiv \neg \LTLdiamond \neg \varphi$. $\LTLdiamond \varphi$ is interpreted as $\varphi$ will \emph{eventually} hold in the future and $\LTLsquare \varphi$ as $\varphi$ holds \emph{globally}.
Operators can be nested, e.g. $\LTLsquare \LTLdiamond \varphi$ states that $\varphi$ has to occur infinitely often.
Linear-time Temporal Logic (LTL)~\cite{pnueliTemporalLogicPrograms1977} is the prototypical temporal logic for expressing requirements of reactive systems.
For example, the following formula describes an arbiter: Given two processes and a shared resource, the formula $\LTLsquare (r_0 \rightarrow \LTLdiamond g_0) \wedge \LTLsquare (r_1 \rightarrow \LTLdiamond g_1) \wedge \LTLsquare \neg (g_0 \wedge g_1)$ describes that whenever a process requests $(r)$ access to a shared resource, it will eventually be granted $(g)$.
Formally, the reactive synthesis problem for LTL is defined over the notion of a strategy as follows: An LTL formula $\varphi$ over atomic propositions $\mathit{AP} = I~\dot\cup~O$ is realizable if there exists a strategy $f: (2^I)^* \rightarrow (2^O)$ that satisfies $\varphi$.
We show the formal syntax and semantics of LTL and the definition of a strategy in the Appendix~\ref{app:ltl}.

\paragraph*{And-Inverter Graphs.} And-Inverter Graphs are directed acyclic graphs that represent reactive systems using three fundamental building blocks: the AND gate, the inverter (NOT gate), and latches, which can store a single bit for one time-step. The graph's edges define the connections between gates, indicating how signals propagate through the circuit. And-Inverter Graphs, especially the AIGER format \cite{biereAIGERAndInverterGraph2007,biereAIGER2011}, are widely used in formal verification and reactive synthesis. 
The AIGER format follows a well-defined specification. The first line contains header information: the maximal variable id, the number of inputs, outputs, latches, and AND gates in the circuit. The circuit's components are following in this order: inputs, latches, outputs, AND-gates, with each component in one line. Each input, AND-gate, and latch defines an even number (variable id) to which other gates and outputs can refer to establish connections between gates. NOT gates are implicitly encoded by the odd version of each number. True and False are encoded by the numbers \texttt{1} and \texttt{0}. We provide an example in Figure~\ref{app:output_fornat} in the Appendix.

\paragraph*{Deep Learning in Formal Methods.}

Deep Learning methods have been successfully applied to various domains in formal methods.
Applications of deep learning methods in symbolic reasoning include 
SAT/SMT solving~\cite{balunovicLearningSolveSMT2018,cameronPredictingPropositionalSatisfiability2020,selsamGuidingHighperformanceSAT2019,selsamLearningSATSolver2019},
temporal logics such as 
generating satisfying traces~\cite{hahnTeachingTemporalLogics2021}, 
reactive synthesis and repair~\cite{coslerIterativeCircuitRepair2023,kretinskyGuessingWinningPolicies2023,schmittNeuralCircuitSynthesis2021}, as well as 
generating symbolic reasoning problems in temporal logics and symbolic mathematics~\cite{kreberGeneratingSymbolicReasoning2023}. 
Mathematical reasoning problems, including integration and differential equations, have been approached with transformers~\cite{lampleDeepLearningSymbolic2020} and through code generation with Large Language Models (LLMs)\cite{droriNeuralNetworkSolves2022}. 
Mathematical reasoning has also been tackled through automatic proof generation~\cite{liIsarStepBenchmarkHighlevel2021}. 
More general applications of deep learning to theorem proving are guiding the proof search with 
clause selection for CNF formulas~\cite{loosDeepNetworkGuided2017} and 
tactic and premise selection/prediction for Coq and HOL light~\cite{bansalHOListEnvironmentMachine2019,bansalLearningReasonLarge2020,huangGamePadLearningEnvironment2019,paliwalGraphRepresentationsHigherorder2020}. 
In contrast to proof guidance, LLMs can be used for end-to-end generation and repair of proofs in Isabelle/HOL~\cite{firstBaldurWholeProofGeneration2023}. LLMs have recently also enabled a step towards autoformalization of unstructured natural language for 
theorem proving~\cite{jiangDraftSketchProve2023,wuAutoformalizationLargeLanguage2022} and 
temporal logic~\cite{coslerNl2specInteractivelyTranslating2023}.
Further, deep learning has had a considerable impact on program verification and synthesis, i.e., for 
termination analysis~\cite{alonUsingGraphNeural2022,giacobbeNeuralTerminationAnalysis2022}, creating 
loop invariants~\cite{peiCanLargeLanguage2023,ryanCLN2INVLearningLoop2019,siLearningLoopInvariants2018} and 
program synthesis/induction~\cite{aletLargescaleBenchmarkFewshot2021,clymoDataGenerationNeural2020,ellisDreamCoderGrowingGeneralizable2023,fijalkowScalingNeuralProgram2022}.

\section{The Neuro-symbolic Portfolio Solver \tool}
\label{sec:neurosynt}
The portfolio solver provides a unified approach to neural and symbolic methods for reactive synthesis. For a seamless integration of the neural method, \tool~relies on model checking (for soundness) and is backed up by symbolic synthesis tools (for completeness).

\subsection{Overview}

\begin{figure}[t]
  \centering
  \includegraphics[width=\textwidth]{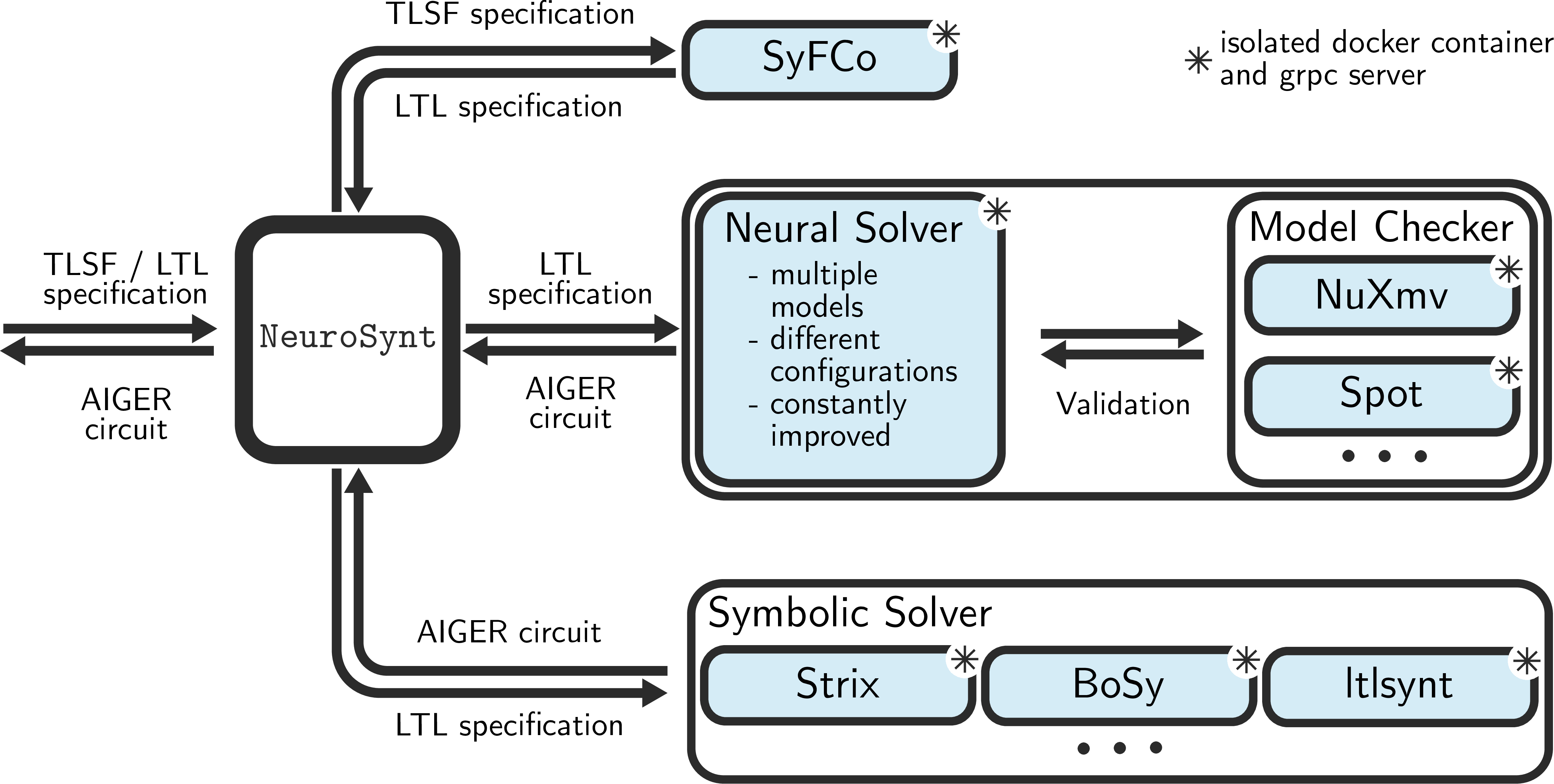}
 \caption{An overview of \tool.}
 \label{fig:overview}
 \end{figure}
We provide an overview of \tool~in the following. Figure~\ref{fig:overview} shows the system design of \tool.
With a single call, a sample is 1) translated from TLSF \cite{jacobsHighlevelLTLSynthesis2016}, the standardized input format for reactive synthesis, to LTL assumptions and guarantees. 2) Fed into the neural solver described in Section \ref{sec:neural_solver} with candidate solutions being verified by a model-checker. This is a feasible approach since LTL model checking is computationally significantly easier than reactive synthesis (PSPACE~\cite{sistlaComplexityPropositionalLinear1985} vs. 2-EXPTIME~\cite{pnueliSynthesisReactiveModule1989a}). 3) A symbolic solver is queried simultaneously with the neural solver. The final result is an implementation in the form of an AIGER~\cite{biereAIGERAndInverterGraph2007} circuit, which is either a verified candidate circuit of the neural solver or the circuit returned by the symbolic solver. Depending on the specification's realizability, the circuit either represents the system implementation (proving realizability) or the environment behavior (proving unrealizability).

All components, neural solver, symbolic solver, and model-checker, are isolated Docker containers. All communication channels between components are defined through a standardized API. Therefore, extending, maintaining, and updating tools are uncoupled from \tool's implementation. Currently integrated are solvers based on the Python library ML2\footnote{\url{https://github.com/reactive-systems/ml2}}, including the neural solver described in Section \ref{sec:neural_solver}, nuXmv \cite{cavadaNuXmvSymbolicModel2014}, NuSMV \cite{cimattiNuSMVOpenSourceTool2002}, Spot \cite{duret-lutzSpotSpot102022}, Strix \cite{meyerStrixExplicitReactive2018}, and BoSy \cite{faymonvilleBoSyExperimentationFramework2017}. We use SyFCo~\cite{jacobsHighlevelLTLSynthesis2016} to convert from TLSF to assumptions and guarantees in LTL.

\subsection{Usage}
\label{ssec:usage}
Since NeuroSynt comprises multiple tools that operate in conjunction during each execution, users must specify arguments to tailor the behavior of these tools. 
We categorize these arguments into tool-specific and general arguments to simplify this process.
General arguments are unrelated to any specific tool and
are passed with the execution command in the command-line interface.

For tool-specific arguments, we use the YAML format \cite{ben-kikiYAMLAinMarkup2021} to create configuration files that encompass the neural engine arguments, the chosen tool for model checking, and symbolic synthesis tasks, along with their respective arguments.
These configuration files facilitate reproducibility and provide a structured way to manage tool-specific settings. An example of a configuration file is demonstrated in Figure~\ref{app:configuration} in the Appendix.

Depending on user choice, \tool~can either wait for all tools to finish/timeout and report all results or return the fastest solution.
We allow the standardized input format TLSF \cite{jacobsHighlevelLTLSynthesis2016} and simple assume-guarantee structured files in LTL (see Figure~\ref{app:input_format} in the Appendix for an example).

\tool~offers two primary execution commands: \emph{benchmark} for solving a dataset of samples and \emph{synthesize} for processing individual samples. For \emph{benchmark}, all results are saved in a CSV file, which can be further analyzed. In all other cases, the result is printed to the command line.
First, we indicate whether the specification was found to be REALIZABLE or UNREALIZABLE, after which we print the system in AIGER format \cite{biereAIGERAndInverterGraph2007} (see Figure~\ref{app:output_fornat} in the Appendix).

\subsection{Implementation and Extensibility}
\label{ssec:implementation}

The central design goal of \tool~is to provide interfaces that are easy to implement when adding and integrating new components.
We first describe the communication interfaces between components. Secondly, we detail some of the messages, and lastly, we describe the options to extend the portfolio solver.

Each solver or model-checker is isolated in a Docker container and communicates with other components through gRPC interfaces.
gRPC is a high-performance open-source framework initially developed by Google for building remote procedure call (RPC) APIs.
Protocol buffers (protobuf) are used as the interface definition language, ensuring programming-language-agnostic interfaces.

\begin{figure}[t]
  \centering
  \includegraphics[width=\textwidth]{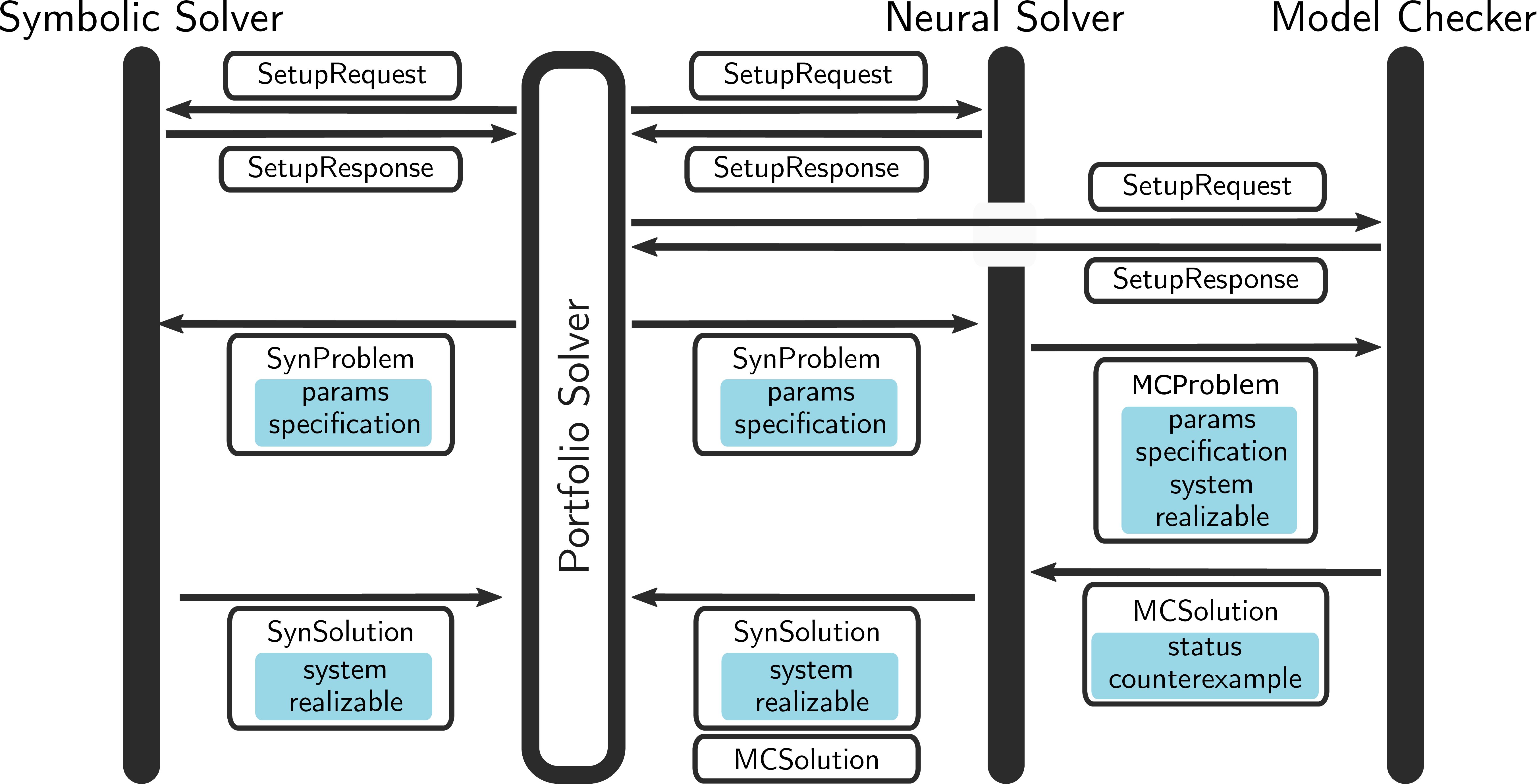}
  \caption{Communication diagram of gRPC calls for a run of \tool, calling the Symbolic Solver and the neural solver, including model-checking.}
  \label{fig:communication}
\end{figure}

In Figure~\ref{fig:communication}, we show the communication through gRPC APIs for the run of \tool~with one specification.
In the first step, each tool is initialized using setup messages, ensuring the components' successful connection.
After setup, a synthesis problem call is sent to the symbolic and neural solver in parallel.
Both solvers eventually report with a synthesis solution.
Before responding, the neural solver makes one or multiple calls to the model checker with candidate solutions, the specification, and the information on whether the specification is suspected to be realizable.
The model checker answers a status and optionally a counterexample.
The neural solver then selects one solution if multiple candidates have been generated and responds to \tool.
The following details the specific protobuf messages that can be exchanged between components.

\paragraph*{SetupRequest and SetupResponse.}
As initialization, the components exchange simple messages through a JSON-like object. 
This message establishes the successful connection and allows the user to provide some tool- but not run-specific arguments. In the case of the neural solver, the model name and other parameters are transmitted to load the model into the memory. The component then responds with a simple success flag or error message.

\paragraph*{SynProblem, SynSolution, and UnsoundSynSolution.}
The \emph{SynProblem} (request) contains an LTL Specification and a set of JSON-like parameters to configure the run- and tool-specific arguments, such as timeout or the number of threads.
The LTL specification is decomposed into guarantees and assumptions, both strings in infix or prefix notation. A \emph{SynSolution} contains the system as the string representation of an AIGER circuit or mealy machine, a status (realizable, unrealizable, error, timeout, nonsucess), the calculation duration, and the tool's name.
No system must be returned if error, timeout, or nonsucces were reported.
The \emph{UnsoundSynSolution} consists of a \emph{SynSolution} and \emph{MCSolution} and is returned by the neural solver.
We show the protobuf definition for the \emph{SynSolution}, \emph{SynProblem}, and the Specification in Figure~\ref{fig:protobuf}. Other definitions can be found in the Appendix in Figure~\ref{app:protobuf}.

\begin{figure}[t]
  \centering
  \begin{lstlisting}[language=protobuf2,style=protobuf]
// An LTL Synthesis solution. Used as response message for Synthesis.
message LTLSynSolution {
  // AIGER circuit. It is allowed to pass no system, e.g. if a timeout 
  // happened. 
  optional AigerCircuit circuit = 1;
  // Shows, whether the specification was found to be realizable or
  // unrealizable. May not be set, e.g. if a timeout happened.
  optional bool realizable = 2;
  // A status that includes useful information about the run.
  LTLSynStatus status = 3;
  // Here additional information should be supplied if the status value
  // requires more details.
  string detailed_status = 4;
  // which tool has created the response.
  Tool tool = 5;
  // How long the tool took to create the result.
  optional google.protobuf.Duration time = 6;
}

message LTLSynProblem {
  // Defines run- and tool-specific parameters. As Map (Dict in Python).
  // Typical examples are threads, timeouts etc. Can be empty.
  map<string, string> parameters = 1;
  // A decomposed specification (assumptions + guarantees).
  DecompLTLSpecification decomp_specification = 2;
}

message DecompLTLSpecification {
  // All input atomic propositions that occur in guarantees or assumptions. 
  repeated string inputs = 1;
  // All output atomic propositions that occur in guarantees or assumptions
  repeated string outputs = 2;
  // A set of guarantees that make up the specifications. All inputs and
  // outputs occurring in any guarantee must be part of input/output.
  repeated LTLFormula guarantees = 3;
  // A set of assumptions that make up the specifications. All inputs and
  // outputs occurring in any guarantee must be part of input/output.
  repeated LTLFormula assumptions = 4;
}
\end{lstlisting}
  \caption{The protobuf definition for a \emph{SynSolution}, \emph{SynProblem},  and decomposed LTL specification. Slightly simplified for easier comprehension. We refer the reader to the artifact and our repository for the full definitions.}
  \label{fig:protobuf}
\end{figure}

\paragraph*{MCProblem and MCSolution.}
A tool can request its candidate solutions to be model-checked by sending an \emph{MCProblem} request.
This message contains a set of JSON-like parameters to configure the run- and tool-specific arguments, an LTL specification (see \emph{SynProblem}), and a system and status (see \emph{SynSolution}).
The \emph{MCSolution} contains the status of the model-checking and, if violating, a counterexample in the form of an error trace and the duration of the computation.
We show the relevant protobuf definitions in Figure~\ref{app:protobuf} in the Appendix.

\tool~can be extended in three major ways.
New neural solvers, symbolic solvers, and model-checking tools can be integrated.
Although not required, we recommend wrapping all components into Docker containers as it helps reproducibility, portability, and isolation, especially when run on high-performance clusters.

\paragraph{Neural Solver.}
The neural solver sits at the core of the portfolio solver, with connections to both the model-checking component and the main portfolio solver.
This component has to support receiving and responding to a \emph{SetupRequest} and a \emph{SetupResponse} for initialization. Furthermore, it should respond to \emph{SynProblem} requests with \emph{UnsoundSynSolution}.
To verify candidate solutions, the neural solver should initiate communication with the model-checking component to verify candidate solutions. Therefore, it should also support sending \emph{MCProblem} requests and receiving \emph{MCSolution} responses.
The neural solver can be independent of the ML2 library if it implements the two communication interfaces mentioned above. It can also be based on the ML2 library, where one could benefit from the existing infrastructure ML2 provides.

\paragraph{Model checking tools.}
A model checker should respond to a \emph{SetupRequest} with a \emph{SetupResponse} and receive the \emph{MCProblem} request, perform model checking and answer with an \emph{MCSolution}.

\paragraph{Symbolic Solver.} New symbolic solvers can be integrated into \tool~by implementing the server side of our generic protocol buffer interface for symbolic solvers.
As for all components, a symbolic solver should implement a setup message (\emph{SetupRequest}, \emph{SetupResponse}).
For a synthesis call, the symbolic solver receives a \emph{SynProblem}, performs the synthesis task, and eventually responds with a \emph{SynSolution}.
At the time of writing, we do not require the output of synthesis tools to be model-checked.
However, one can implement the interface to the model checking component to increase the trust in the output of new symbolic approaches.

\section{The Neural Solver}
\label{sec:neural_solver}

The neural solver is at the heart of the portfolio solver and is developed jointly with \tool.
We report on the methodology of the neural solver, including architecture, datasets, data generation, training, and evaluation.
We clearly distinguish between previous work \cite{schmittNeuralCircuitSynthesis2021}, introducing a neural approach for reactive synthesis, and improvements that are integrated into \tool, leading to the significantly increased performance on the SYNTCOMP benchmarks.

\subsection{Data and Data Generation Improvement}

We significantly improved the training data generation compared to previous work.
While the basic algorithm is taken from \cite{schmittNeuralCircuitSynthesis2021}, we scale the size of the training samples, tweak the data generation parameters to fit the larger samples, and combine multiple data generation strategies to lift previous limitations.

We aim for a dataset containing specifications (assumptions and guarantees) and circuits.
Depending on the specification, the circuit is either a winning strategy for the system (realizable) or a winning strategy for the environment (unrealizable).
For each sample, we use an additional token to show whether the system is realizable or unrealizable.
The dataset is for supervised training, with the specification being the input and the circuit, along with the realizability token being the model's target.

In total, we combine three datasets and generation techniques.
For the first two datasets, we utilize the generation method from \cite{schmittNeuralCircuitSynthesis2021} with 1) the minor tweak of having a variable number of inputs and outputs (up to five) in the circuit instead of exactly five (denoted \texttt{previous}), and 2) extensions to handle a larger number of patterns, larger patterns, and patterns with more atomic propositions (denoted \texttt{new}).
The third dataset is a data augmentation method based on the result of \texttt{new} (denoted \texttt{augmented}).

\paragraph*{Data Generation.}
We first report on the data generation algorithm for \texttt{previous} and \texttt{new}.
The data generation has two major steps. 
In the first step, we mined LTL formula patterns that are common in research and practice.
Considering formula patterns is a widespread idea, e.g.,~\cite{dwyerPatternsPropertySpecifications1999}.
We collect patterns from $1075$ (previously $346$) benchmarks from the LTL synthesis track of SYNTCOMP 2022.
We extract a list of $627$ assumption patterns and $7948$ guarantee patterns.
An assumption restricts the environment, and a guarantee defines the implementation's behavior.
To fit the model requirements, we filtered out LTL formulas with more than $15$ inputs and $15$ outputs (previously $5$).
Additionally, we filter out specifications with an abstract syntax tree (AST) size greater than $30$ (formerly $25$), resulting in $519$ (formerly $157$) assumption patterns and $6841$ (previously $1942$) guarantee patterns.
In the second step, we constructed synthesis specifications by combining the mined patterns.
For each specification, we alternate between sampling guarantees until the specification becomes unrealizable, and sampling assumptions until the specification becomes realizable.
Whether we aim for a realizable or unrealizable specification, we either collect the last successfully mined specification (realizable) or the second-to-last mined specification (unrealizable).
We aim for an even split between realizable and unrealizable specifications.
To handle more atomic propositions while reducing patterns that do not share atomic propositions, we now favor atomic propositions present in the already constructed part of the specification with a bias of 4 when instantiating the patterns.
We continue this process until we reach one of the following stopping criteria:

\begin{itemize}
  \item[a)] the specification has the maximal number of guarantees (10),
  \item[b)] the specification has the maximal number of assumptions (3),
  \item[c)] the synthesis tool timed out (120s timeout), or
  \item[d)] no suitable assumption was found after 7 (formerly 5) attempts.
\end{itemize}

To ensure an even distribution of challenging instances, we filter AIGER circuits exceeding a maximum variable index of 60 and only allow a certain amount ($20\%$) of circuits with the same number of AND gates.

\paragraph*{Data Augmentation.}
We augment the dataset \texttt{new} as a third approach to artificially force larger properties for a share of the final dataset.
For each specification in \texttt{new}, we combine multiple patterns into one property until we reach an AST size of $30$.
Having longer properties in the training dataset leads to better generalization to even larger properties.
Compared to \texttt{new}, the augmented dataset has an average of $3$ guarantees instead of $5.6$, with an average size of $22.9$ per guarantee instead of $12.3$.

\begin{figure}[t]
  \centering
  \includegraphics[width=\textwidth]{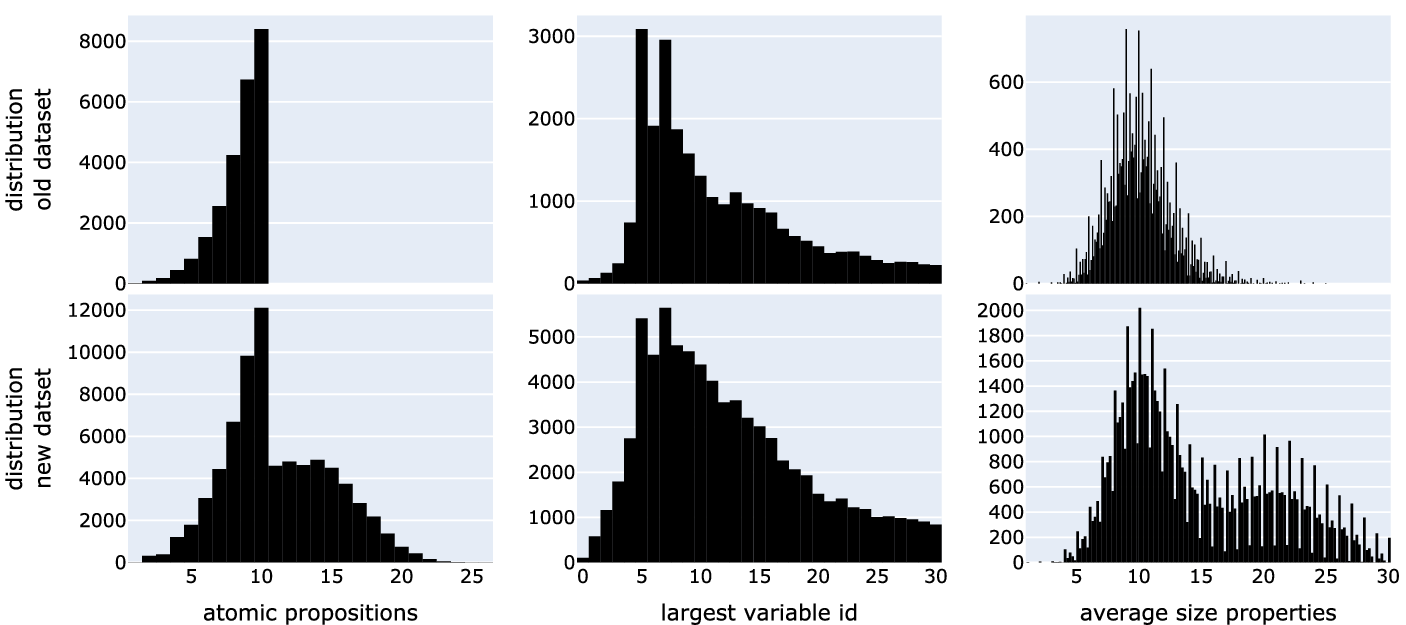}
  \caption{Previous dataset\cite{schmittNeuralCircuitSynthesis2021}, compared with the new final dataset. Comparing the number of atomic propositions in a sample, the largest variable id in the AIGER circuit, and the average size of properties.}
  \label{fig:dataset}
\end{figure}

\paragraph*{Final Dataset.}
All three resulting datasets are combined into a single dataset, consisting of $600\,000$ training samples and $75\,000$ validation and test samples. Figure~\ref{fig:dataset}, shows the key differences in features of the new final dataset compared to the previous dataset~\cite{schmittNeuralCircuitSynthesis2021}.
While the previous dataset used only up to $5$ inputs and outputs in the specification, we now have up to $15$ inputs and outputs, leading to up to $25$ atomic propositions in a specification.
We also have slightly more latches in the new dataset ($1.23$ instead of previously $1.16$).
Note that the same version and configuration of Strix \cite{meyerStrixExplicitReactive2018} was used in both approaches.
The most apparent distinction to previous datasets\cite{schmittNeuralCircuitSynthesis2021} is in the size of the properties, where we clearly see the effects of the data augmentation process.

\subsection{Architecture \& Training}

\paragraph*{Transformer Architecture.}
The core of the neural solver implemented in \linebreak[4]\tool~is a Transformer neural network~\cite{vaswaniAttentionAllYou2017}.
The vanilla Transformer architecture follows a basic encoder-decoder structure.
The encoder constructs a hidden embedding $z_i$ for each input embedding $x_i$ of the input sequence $x = (x_0, \ldots, x_n)$ in parallel.
An embedding is a mapping from plain input, for example, words or characters, to a high dimensional vector, for which learning algorithms and toolkits exist, e.g., word2vec~\cite{mikolovEfficientEstimationWord2013}.
Given the encoders output $z = (z_0, \ldots, z_k)$, the decoder generates a sequence of output embeddings $y = (y_0, \ldots , y_m)$ autoregressively.
Since the transformer architecture contains no recurrence nor convolution, we apply a tree positional encoding~\cite{shivNovelPositionalEncodings2019c}.

\begin{figure}[t]
  \centering
  \includegraphics[width=\textwidth]{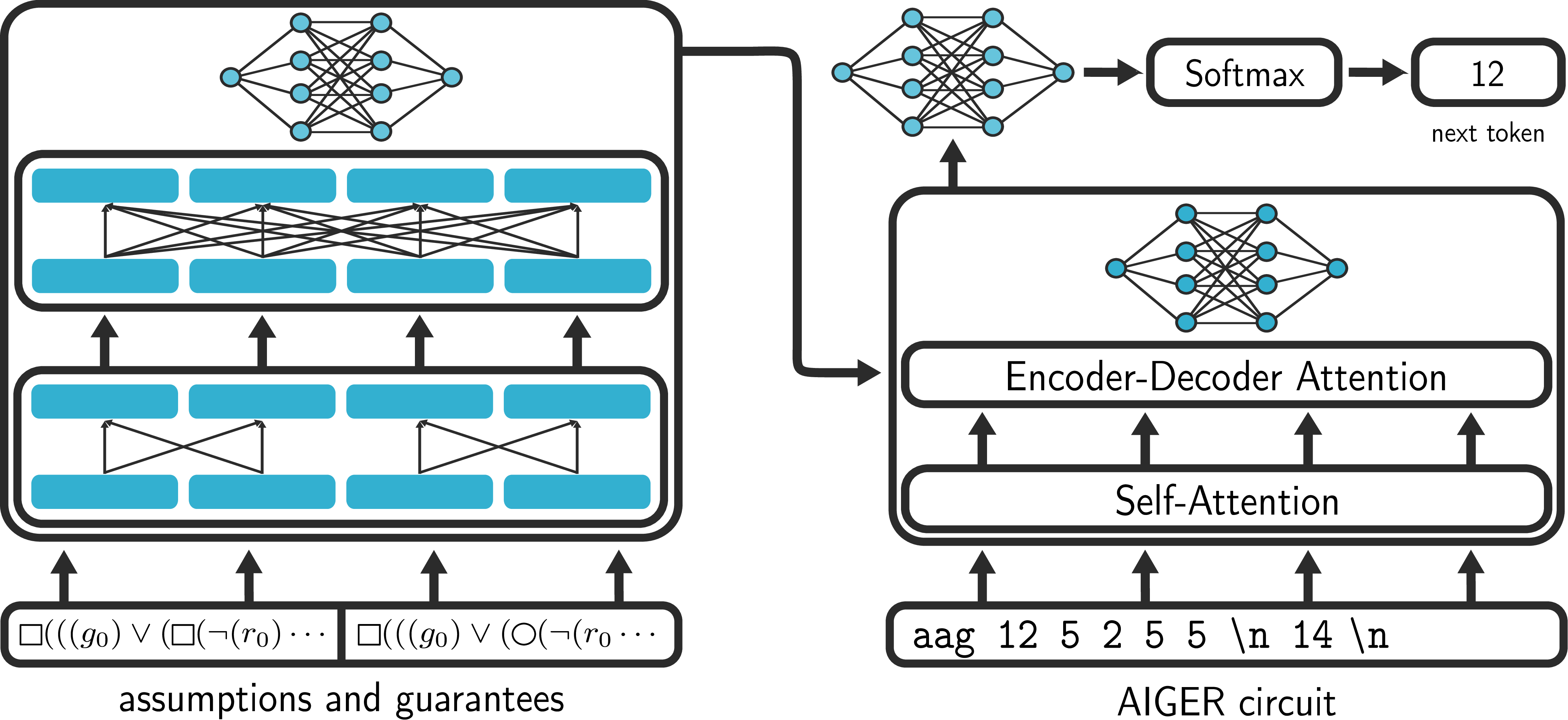}
  \caption{Schematic view of the Hierarchical Transformer, with illustrated inputs/outputs of the reactive synthesis application. The encoder shows the hierarchical self-attention with separation into local and global layers. For simplicity, we show one local and global layer and only two assumptions and guarantees with two tokens each.}
  \label{fig:transformer} 
\end{figure}

The main idea of the Transformer is a self-attention mechanism to compute a score for each pair of input elements, representing which positions in the sequence should be considered the most when computing the hidden embeddings.
For each input embedding $x_i$, we compute 1) a query vector $q_i$, 2) a key vector $k_i$, and 3) a value vector $v_i$ by multiplying $x_i$ with weight matrices $W_k$, $W_v$, and $W_q$, which are learned during the training process.
The embeddings can be calculated simultaneously using matrix operations~\cite{vaswaniAttentionAllYou2017}. Specifically, let
$Q, K, V$ be the matrices obtained by multiplying the input vector X consisting of all
$x_i$ with the weight matrices $W_k$, $W_v$, and $W_q$:
$\mathit{Attention}(Q, K, V) = \mathit{softmax}(\frac{QK^T}{\sqrt{d_k}})V$, with $d_k$ being the model's dimension. For details, we refer the interested reader to~\cite{vaswaniAttentionAllYou2017}.
The Transformer variation used in this paper is a so-called hierarchical Transformer~\cite{liIsarStepBenchmarkHighlevel2021}, separating the encoder self-attention into local and global layers. Local layers embed assumptions and guarantees individually and invariant against their order. Global layers calculate the self-attention across all assumptions and all guarantees. We show an illustration in Figure~\ref{fig:transformer}.

\begin{figure}[t]
  \centering
  \includegraphics[width=\textwidth]{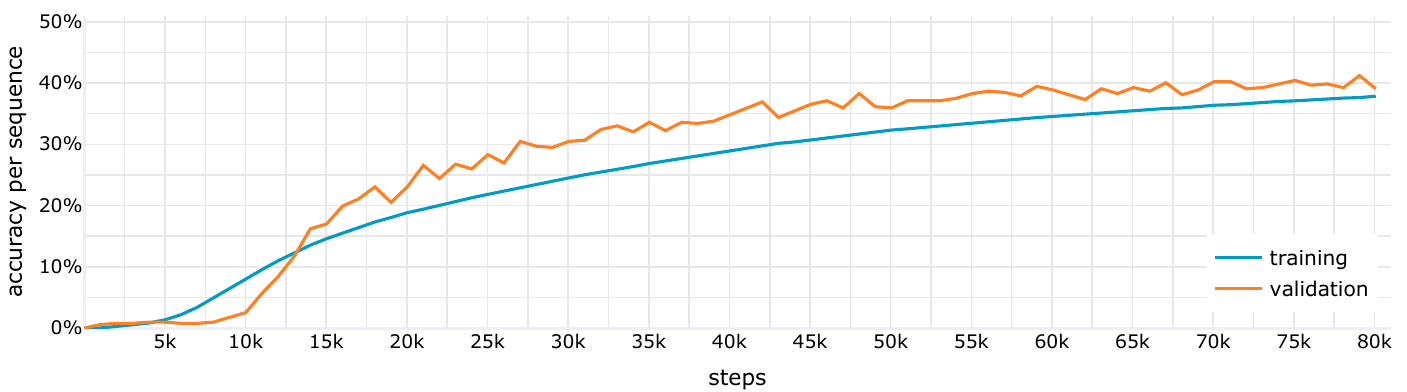}
  \caption{Accuracy per sequence during training. Measured on training and validation data.}
  \label{app:training}
\end{figure}

\paragraph*{Model Hyperparameter \& Training.}
We train our model on the $600\,000$ samples from our training dataset for $80\,000$ steps with early stopping and a batch size of $512$. We show a plot of the accuracy per sequence in Figure~\ref{app:training}.
We train data parallel on two Nvidia A100 40GB from a Nvidia DGX A100 system, which takes approximately $10$ hours.
We use the Adam optimizer \cite{kingmaAdamMethodStochastic2015} with $\beta_1 = 0.9$, $\beta_2 = 0.98$ and $\epsilon = 10^{-9}$. 
We use learning rate scheduling as proposed in \cite{vaswaniAttentionAllYou2017} with $4000$ warmup-steps.
Our model consists of $4$ local, $4$ global, and $8$ decoder layers, each having $4$ heads.
All feed-forward networks have $1024$ nodes to which we apply a dropout of $0.2$. Our model has a total size of $14\,791\,748$ parameters. Input and output tokens have an embedding of size $256$. The maximum input and target lengths are set according to the training data with at most $12$ properties, a maximum AST size of $32$ per property for the specification, and a maximum circuit length of 128 tokens after encoding. 

We show that our model significantly improved compared to previous work \cite{schmittNeuralCircuitSynthesis2021} by reimplementing and adapting the previous model to evaluate the 2022 SYNTCOMP benchmarks. With $21.8\%$, the new model improved by $13$ percentage points to $34.83\%$. We explore more details of the evaluation of the model in Section~\ref{ssec:generalization}.

\section{Experiments \& Benchmarks}
\label{sec:experiments}

We split our experiments into two segments. In Section~\ref{ssec:generalization}, we first perform generalization experiments on the integrated neural solver. The neural solver can generalize on its training distribution but also to more complex instances, longer specifications, and out-of-distribution instances, which we show using the datasets \texttt{test}, \texttt{large}, \texttt{timeouts}, and \texttt{syntcomp}.

Secondly, in Section~\ref{ssec:comparisons}, we evaluate the performance of the \tool~framework on the SYNTCOMP 2022 benchmarks.
To this end, we use \tool~to compare the performance of the neural solver against multiple symbolic solvers and highlight efficiency gains and enhancements that arise from the combination of both methodologies.
We show that the combined effort of neural and symbolic solvers leads to a performance gain that symbolic solvers alone could not achieve.

The evaluation is performed on a GPU cluster (1 Nvidia DGX A100 40GB, AMD EPYC 7F32 @ 1.8GHz base, 3.7GHz max, 8 cores + 8 SMT cores, 256GB RAM), on a CPU cluster (Intel Xeon E7-8867 v3 @ 2.50GHz, 64 cores + 64 HT cores, and 1536 GB RAM) and additionally did some early experiments on an Apple M1 Max (64GB memory, $10$ cores, 32 neural cores).

Similar to different configurations of symbolic solvers, we have multiple models with slightly different performances. This paper reports the results of the model that performed best on the SYNTCOMP benchmark. Whenever we consider additional models, we mention that explicitly.

\subsection{Generalization}
\label{ssec:generalization}

We analyze the generalization capabilities of the model in the neural solver in four ways.
Firstly on our \texttt{test} set, secondly on samples that are significantly larger than seen during training (\texttt{large}), thirdly samples that are arguably more difficult than training samples, and fourthly on out-of-distribution samples (\texttt{syntcomp}). Here, we consider instances that are not from the same data generation algorithm out-of-distribution samples.
Results on these datasets are in Table~\ref*{table:results}.

\begin{table}[b]
  \centering
  \setlength{\tabcolsep}{0.7em}
  \caption{Performance of the neural model on different datasets.}
  \label{table:results}
  \begin{tabular}{@{}lllllll@{}}
  \toprule
                                                               & test     & large    & timeouts & \begin{tabular}[c]{@{}l@{}}syntcomp-\\ small\end{tabular} & \begin{tabular}[c]{@{}l@{}}syntcomp-\\ large\end{tabular} & \begin{tabular}[c]{@{}l@{}}syntcomp-\\ full\end{tabular} \\ \midrule
  \begin{tabular}[c]{@{}l@{}}syntactic\\ accuracy\end{tabular} & $38.6\%$ & $10.2\%$ & -        & -                                                         & -                                                         & -                                                        \\
  \begin{tabular}[c]{@{}l@{}}semantic\\ accuracy\end{tabular}  & $84.2\%$ & $57.7\%$ & $33\%$   & $65.8\%$                                                  & $54.5\% $                                                 & $34.83\% $                                               \\ \bottomrule
  \end{tabular}
  \end{table}

\paragraph*{Generalization on \texttt{test} and \texttt{large}.} 
On the datasets \texttt{test} and \texttt{large}, additionally to measuring correct solutions (semantic accuracy, $84.2\%$), we collect how many solutions are syntactically identical to the solution from our data generation algorithm ($38.6\%$).
The large difference of $45.6$ percentage points on our \texttt{test} dataset indicates that the neural solver generalizes to the semantics of the synthesis problem instead of learning the particularities of the data generator.

The dataset \texttt{large} consists of larger samples than seen during training.
Samples in \texttt{large} have at least $10$, on average $14.5$ properties, and the largest property in each sample has an AST of $37.9$ on average. In contrast, training samples have $5.3$ properties on average, with the largest property having an AST of $22.2$ on average.

\begin{figure}[t]
  \centering
  \includegraphics[width=\textwidth]{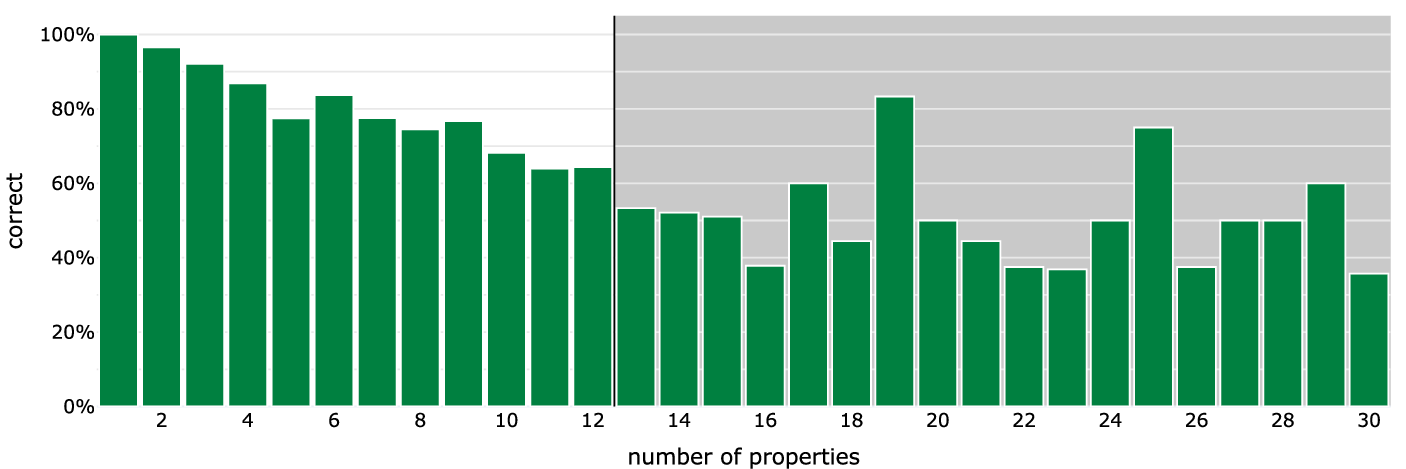}
  \includegraphics[width=\textwidth]{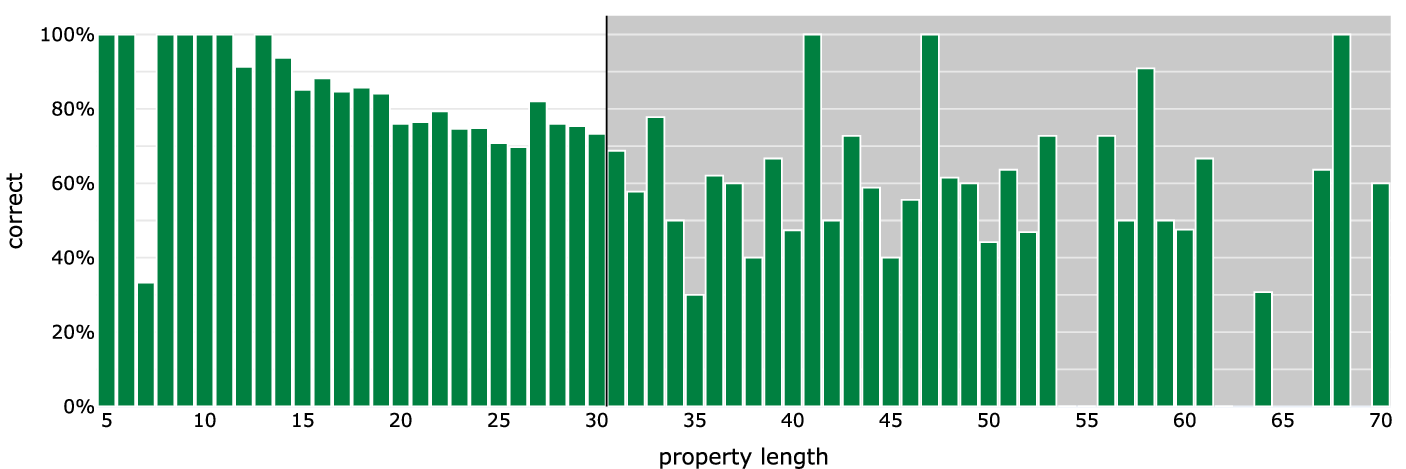}
  \caption{Share of correct solutions on the joint dataset of \texttt{large} and \texttt{test} over the number of properties in a sample and the size of the largest property in each sample. A darker background indicates sizes larger than seen during training.}
  \label{evaluation:share_large}
\end{figure}

For a more detailed analysis, we join datasets \texttt{test} and \texttt{large} and plot the share of correct solutions partitioned by the number of properties as well as the size of the largest property in each sample in Figure~\ref{evaluation:share_large}.
The largest property seen during training is $30$, and the largest number of properties per specification is $12$.
While we see a decrease in performance for larger samples, there is no clear drop after $12$ or $30$, respectively, which indicates generalization with the number of properties and the length of the properties.
Note that results from larger sizes naturally have less significance as fewer samples per bucket exist. We refer to Figure~\ref*{app:dist_large} in the Appendix for the total count of samples in each displayed bucket.

\paragraph*{Generalization on \texttt{timeouts}.}
The dataset \texttt{timeouts} consists of samples on which Strix timed out after $120s$ during our data generation.
Therefore, such samples can be seen as significantly harder, while not larger than samples in the training data.
We achieve $33\%$ correct solutions on this dataset, showing that our model generalizes from the training data to more challenging specifications and solutions that could not have not been solved by Strix during the data generation.
This experiment prognosticates the potential of combining neural methods with symbolic methods.

\paragraph*{Generalization on out-of-distribution dataset.}
While \texttt{large} and \texttt{timeouts} were generated with the same data generation approach as the training data, \linebreak[4]\texttt{syntcomp-full} consists of all $1075$ real-world specifications collected in the SYNTCOMP benchmark, on which the neural solver achieves $34.83\%$ accuracy (see Table~\ref{table:results}).
\texttt{syntcomp-large} contains all such samples that are in the size of our evaluation constraints (i.e. max $30$ properties, max AST size of $70$ per property, $54.5\%$ accuracy).
\texttt{syntcomp-small} contains only such samples that are in the training data size (i.e., max $12$ properties, max AST size of $30$ per property, $65.8\%$ accuracy).
We see a remarkable generalization to out-of-distribution samples with an accuracy of $64.8\%$ on \texttt{syntcomp-small}.
We additionally observe generalization on specification size that we also see on the \texttt{large} dataset.
We refer to Figure~\ref{app:share_syntcomp} in the Appendix for more details.

\subsection{Comparisons and Advantages of Combination}
\label{ssec:comparisons}

We demonstrate the advantage of \tool~by comparing the neural solver to the performance of multiple symbolic solvers: Strix~\cite{meyerStrixExplicitReactive2018}, the current state-of-the-art, BoSy~\cite{faymonvilleBoSyExperimentationFramework2017}, a bounded synthesis method, and additionally rely on the results of SYNTCOMP 2022 (ltlsynt~\cite{renkinImprovementsLtlsynt2022}, Otus~\cite{abrahamSymbolicLTLReactive2021}, sdf~\cite{khalimovGamebasedBoundedSynthesis}).
Whenever we write SYNTCOMP in this paper, we refer to the 2022 iteration.

We initiate the evaluation by comparing the neural solver and \tool~to the specified symbolic tools, illustrating the number of problems that can be solved within a specific time frame. Then we dive into details on instances that could only be solved by \tool~and no other symbolic solver (\emph{novel solves}), show details on the \emph{time-to-solution} differences between the solvers, and lastly, look at \emph{circuit sizes} of their respective solutions.\\

\begin{figure}[t]
  \includegraphics[width=\textwidth]{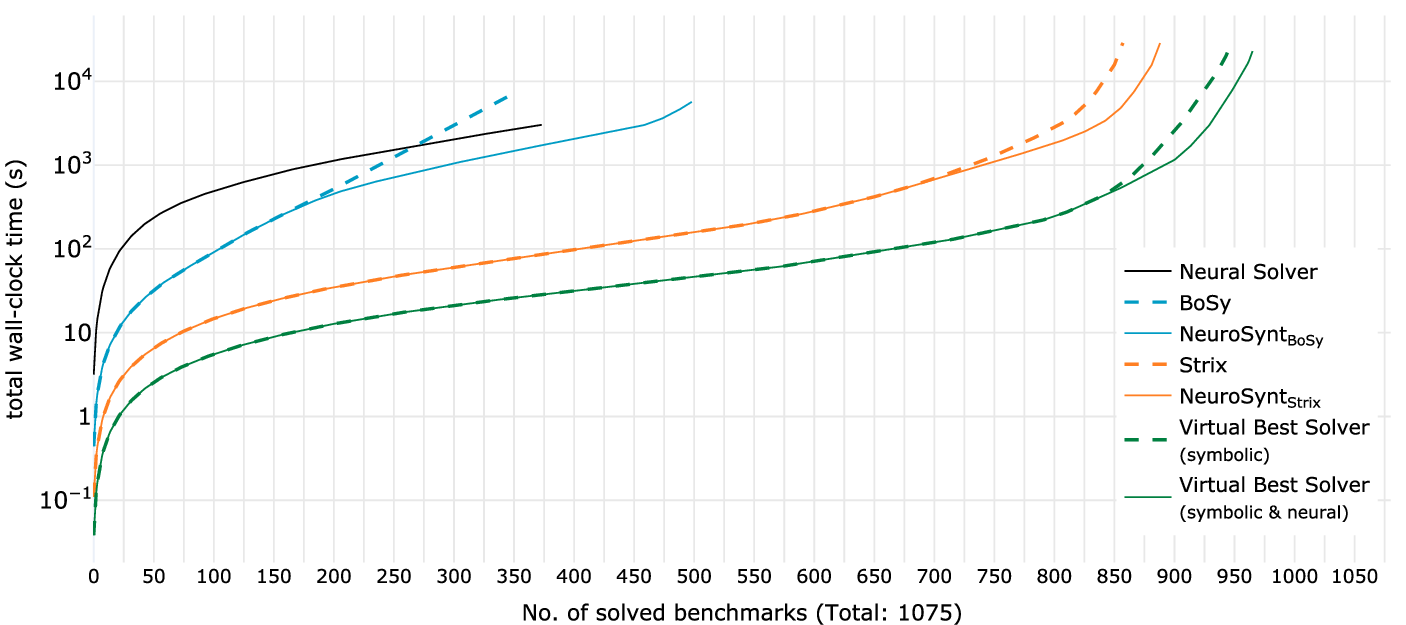}
  \caption{A cactus plot showing the number of solved samples vs. accumulated wall-clock time. Each sample per solver is a dot on the respective line. The lower and further right a line, the better the solver. We compare the neural solver, Strix and BoSy alone, \tool$_\text{BoSy}$ which couples BoSy and the neural solver and \tool$_\text{Strix}$ which couples the neural solver and Strix. Further, we virtually combine the results of all tools and all configurations from the SYNTCOMP to a virtual best solver and compare that with the evaluations of multiple neural models.}
  \label{evaluation:cactus}
\end{figure}

In Figure~\ref{evaluation:cactus}, we display the performance of the neural solver, the performance of several symbolic tools and the performance of \tool~that unites the neural solver with a symbolic solver. We additionally show a Virtual Best Solver (VBS) of all SYNTCOMP 2022 results without and including the neural solver. We further report what the previously published neural reactive synthesis approach \cite{schmittNeuralCircuitSynthesis2021} would have achieved if it had been integrated into the portfolio solver.
With $374$ solved instances, the neural solver alone can already solve more samples than BoSy ($347$) with $120s$ timeout on the CPU cluster.
Its true advantage becomes evident when combining the neural solver with symbolic solvers.
\tool$_\text{BoSy}$ solves $152$ (previous: $59$) samples more than BoSy alone, which is $20.8\%$ of the samples that BoSy could not solve.
Similarly, \tool$_\text{Strix}$ solves $31$ (previous: $2$) samples more than Strix alone ($1h$ timeout on the CPU cluster), which is $14.2\%$ of the samples that Strix could not solve in $1h$.
To show the full potential of \tool, we combined all results from the SYNTCOMP and our experiments with BoSy and Strix.
All symbolic solvers combined were able to solve $945$ instances of the total of $1075$.
Adding the neural solver of \tool~to the virtual best solver, we solve an additional $20$ (previous: $0$) samples exclusively that no other tool tested did solve (\emph{novel solves}). This is $15.4\%$ of the samples that none of the symbolic tools could solve in $1h$.
No other tool in SYNTCOMP 2022 except the state-of-the-art Strix, solved more samples that no other tool could solve. We refer to the appendix~\ref{app:syntcomp-performance-exclusive} for exact numbers.
This signifies that even for specifications that pose computational challenges to symbolic synthesis tools, there exist patterns that a neural network can recognize and exploit post-training.

\paragraph*{Novel Solves.}
Of the $20$ novel solves, $6$ instances are parameterized versions of full arbiters with $3$ processes.
This version of the full arbiter is unrealizable as the specification additionally enforces two grants to hold at the same time step (step $11$ to step $16$ respectively).
These are the largest parameterizations of this problem class in the SYNTCOMP dataset.
Similarly, $11$ instances are full arbiter with $3$ processes, where two grants are enforced simultaneously (step $6$ to $16$, respectively).
These parameterizations are also the largest parameterizations of this problem class in the SYNTCOMP dataset.
One instance is a full arbiter with $6$ processes and the requirement of two grants to hold at \emph{any} time step.
Finally, we have one instance of a load balancer with $6$ grants and the additional unrealizable requirement of two grants at time step $5$.
This is also the largest parameterization of this problem class in the SYNTCOMP dataset.
In the Appendix, Figure~\ref{fig:parameterized_arbiter}, we give an example of such samples.

\paragraph*{Time To Solution.}
For experiments with \tool, we record the wall-clock time of the neural solver, the symbolic solver, and the model checker.
The neural solver (including model checking) is fastest on the GPU cluster, with $8.6s$ and a standard deviation of only $3.3s$.
The time for model-checking using NuXmv is almost negligible, with $0.35s$ on average per sample.
The low standard deviation highlights the advantage of the neural solver, as the time does not depend on the complexity of the specification.
Strix with a timeout of $1h$ on the CPU cluster takes $33.4s$ on average, with a standard deviation of $185.3s$.
We find that the neural solver can also be run on CPU-only hardware (CPU cluster) with an average of $79.4s$ and on hybrid desktop hardware such as the Apple M1 Max with an average of $17.8s$. For an extensive overview over the experiments with different timeouts, we refer the reader to the appendix~\ref{app:times}.

\begin{figure}[t]
  \centering
  \includegraphics[width=\textwidth]{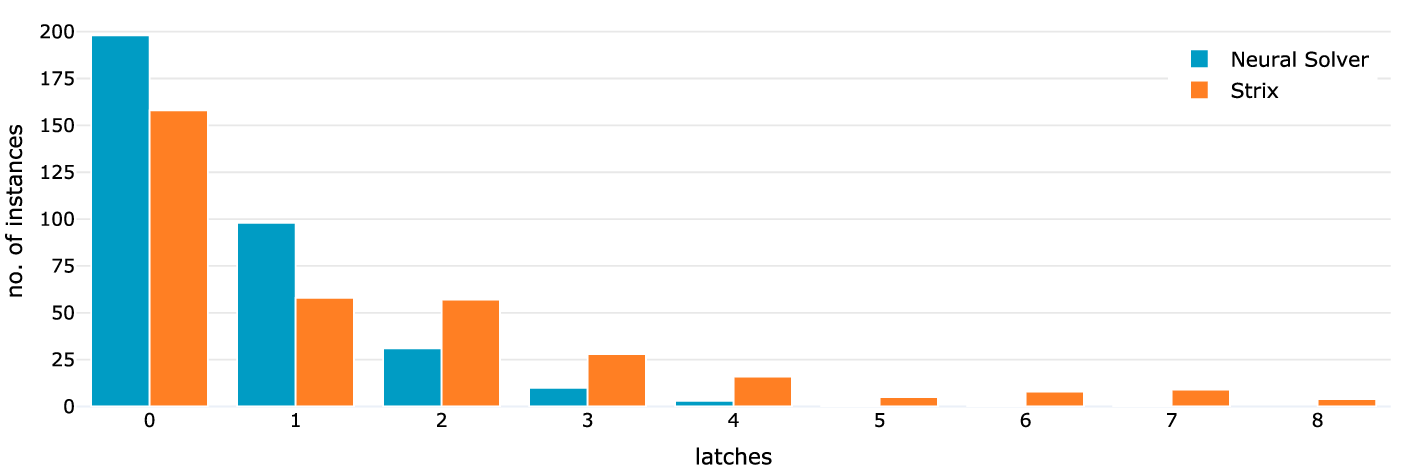}
  \caption{No. of latches per instance. On instances that the neural solver and Strix commonly solved}
  \label{evaluation:latches_strix}
\end{figure}

\paragraph*{Circuit Sizes.}
We find that on instances where the neural solver and the symbolic solver both found a solution, the solution by the neural solver is often smaller than the symbolic solver's.
This holds for BoSy and Strix, but also for all other tools in SYNTCOMP (on the realizable fraction).
On samples solved by Strix and the neural solver, the solutions by the neural solver have $54.9\%$ fewer latches than those by Strix.
In Figure~\ref{evaluation:latches_strix}, we show the distribution of latches for this comparison.
For more details, we refer the reader to the Appendix~\ref{app:sizes}.

\section{Conclusion}

We introduced \tool, a neuro-symbolic portfolio solver for reactive synthesis.
At the core of the portfolio solver lies an integrated neural solver that computes candidate implementations, which are automatically checked by model-checking tools.
We reported on the neural solver's methodology and training and the API framework's implementation to isolate components. The open-source implementation of \tool~provides an interface in which new neural and symbolic approaches alike can be seamlessly integrated.

Our experiments on the generalization capabilities of the Transformer show the ability to generalize to larger specifications, more difficult specifications, and out-of-distribution specifications. The relatively small size of the underlying Transformer neural network suggests that the overall performance of neural solvers can be further increased.

We evaluated the overall performance of \tool, enhancing the state-of-the-art in reactive synthesis with the integrated neural solver contributing novel solves in the SYNTCOMP 2022 benchmark.
With the almost constant evaluation time of the neural solver, the portfolio solver is often faster than previous approaches.
Furthermore, the integrated neural solver yields smaller implementations than state-of-the-art symbolic tools, including Strix and BoSy.

\section{Data Availability Statement}

\tool~is published open-source on GitHub (\url{https://github.com/reactive-systems/neurosynt}). All data, models, and experiments supporting this paper's results are publicly available. A digital artifact is available at (\url{https://doi.org/10.5281/zenodo.10046523}).

%
%

\pagebreak

\bibliographystyle{splncs04}
\bibliography{mybibliography}

\appendix

\section{Linear-time Temporal Logic (LTL)}
\label{app:ltl}
For a given set of atomic propositions $\LTLAP$, the syntax of LTL formulas over $\LTLAP$ is defined as:
\begin{align*}
	\varphi, \psi \Coloneqq \LTLtrue~|~a~|~\neg \varphi~|~\varphi \land \psi~|~\LTLcircle \varphi~|~\varphi \LTLuntil \psi \enspace ,
\end{align*}

\noindent where $\LTLtrue$ is the Boolean constant, $a \in \LTLAP$, $\neg$ and $\land$ are the Boolean connectives and $\LTLcircle$ and $\LTLuntil$ are temporal operators. We refer to $\LTLcircle$ as the \textit{next} operator and to $\LTLuntil$ as the \textit{until} operator.
Other Boolean connectives can be derived. Further, we can derive temporal modalities such as \textit{eventually} $\LTLdiamond \varphi \coloneqq \LTLtrue \LTLuntil \varphi$ and \textit{globally} $\LTLsquare \varphi \coloneqq \neg \LTLdiamond \neg \varphi$. For a given set of atomic propositions $\LTLAP$, the semantics of an LTL formula over $\LTLAP$ is defined with respect to the set of infinity words over the alphabet $2^{\LTLAP}$ denoted by $\left(2^{\LTLAP}\right)^\omega$. The semantics of an LTL formula $\varphi$ is defined as the language $Words(\varphi) = \{\sigma \in \left(2^{\LTLAP}\right)^\omega \mid \sigma \models \varphi\}$ where $\models$ is the smallest relation satisfying the following properties:
\begin{align*}
	\sigma &\models \LTLtrue\\
	\sigma &\models a & &\mathrm{iff}~ a \in A_0\\
	\sigma &\models \neg \varphi & &\mathrm{iff}~ \sigma \not \models \varphi\\
	\sigma &\models \varphi \land \psi & &\mathrm{iff}~ \sigma \models \varphi ~\mathrm{and}~ \sigma \models \psi\\
	\sigma &\models \LTLcircle \varphi & &\mathrm{iff}~ \sigma[1\ldots] \models \varphi\\
	\sigma &\models \varphi \LTLuntil \psi & &\mathrm{iff}~ \exists j \geq 0.~ \sigma[j\ldots] \models \psi ~\mathrm{and}~ \forall 0 \leq i < j.~ \sigma[i\ldots] \models \varphi \enspace ,
\end{align*}

\noindent where $\sigma = A_0 A_1 \ldots \in (2^{\LTLAP})^\omega$ and $\sigma[i\ldots] = A_i A_{i+1} \ldots$ denotes the suffix of $\sigma$ starting at $i$.

A strategy $f: (2^I)^* \rightarrow 2^O$ maps sequences of input valuations $2^I$ to an output valuation $2^O$.
The behavior of a strategy $f$ is characterized by an infinite tree, called computation tree, that branches by the valuations of $I$ and whose nodes $w \in (2^I)^*$ are labeled with the strategic choice $f(w)$.
For an infinite word $w = w_0 w_1 w_2 \ldots \in (2^I)^*$, the corresponding trace is defined as
$(f(\epsilon) \cup w_0) (f(w_0) \cup w_1) (f(w_1) \cup w_2) \ldots \in (2^{I \dot \cup O})^\omega$.

\section{Usage of NeuroSynt}

We provide an example for an assume-guarantee style input file in Figure~\ref{app:input_format}. It can be used as an alternative to the standardized TLSF~\cite{jacobsHighlevelLTLSynthesis2016} format. Figure~\ref{app:output_fornat} provides the output that \tool~produces for this specification. In Figure~\ref{app:configuration}, we give an example of the configuration file for \tool, running Strix as a symbolic solver, NuXmv as a model checker, and our neural solver.

\begin{figure}[!ht]
  \scriptsize
  \begin{lstlisting}[language=json]
{
  "semantics": "mealy",
  "inputs": ["r_0", "r_1"],
  "outputs": ["g_0", "g_1"],
  "assumptions": [],
  "guarantees": ["(G ((! (g_0)) | (! (g_1))))",
                 "(G ((r_0) -> (F (g_0))))",
                 "(G ((r_1) -> (F (g_1))))"]
}\end{lstlisting}
\caption{Example of the JSON input file of a simple 2-bit-arbiter}
\label{app:input_format}
\end{figure}

\begin{figure}[!ht]
  \scriptsize
  \begin{lstlisting}[language=json]
REALIZABLE
aag 3 2 1 2 0
2
4
6 7
7
6
i0 r_0
i1 r_1
l0 l0
o0 g_0
o1 g_1
      \end{lstlisting}
      \caption{Example of the AIGER output for a simple 2-bit arbiter}
      \label{app:output_fornat}
\end{figure}

\begin{figure}[!ht]
  \scriptsize
  \begin{lstlisting}[language=json]
symbolic_solver:
  tool: strix
  tool_args:
    "timeout": 120
    "--threads": 4
    "--minimize": ""
    "--auto": ""
  service_args:
    "mem_limit": "2g"
    "start_containerized_service": True
model_checker:
  tool: nuxmv
  tool_args:
    "timeout": 10
  service_args:
    "mem_limit": "2g"
    "start_containerized_service": True
neural_solver:
  tool: ml2solver
  service_args:
    "nvidia_gpus": False
    "mem_limit": "100g"
    "start_containerized_service": True
    "start_service": False
  tool_setup_args:
    "batch_size": 1
    "alpha": 0.5
    "num_properties": 40
    "length_properties": 70
    "beam_size": 32
    "check_setup": True
    "model": "ht-50"
      \end{lstlisting}
      \caption{Example of the configuration file, running Strix as a symbolic solver, NuXmv as a model checker, and our neural solver.}
      \label{app:configuration}
      \vspace*{-2em}
\end{figure}

\clearpage
\section{Protocol Buffers}
\label{app:proto_buffer}

Figure~\ref{app:protobuf} shows some of our protocol buffer interfaces. We omitted some messages and definitions for simplification. We refer the interested reader to the artifact.
\begin{figure}[!ht]
    \centering
    \begin{lstlisting}[language=protobuf2,style=protobuf]

message LTLFormula {
  // Represents an LTL formula as string.
  string formula = 1;
  // The notation in which the formula is serialized. Infix is default.
  string notation = 2;
}

message UnsoundLTLSynSolution {
  LTLSynSolution synthesis_solution = 1;
  // A model-checking result can optionally be included in the response.
  optional LTLMCSolution model_checking_solution = 2;
  // which tool has created the response.
  Tool tool = 3;
  // How long the tool took to create the result.
  optional google.protobuf.Duration time = 4;
}

message LTLMCProblem {
  // Defines run- and tool-specific parameters. As Map (Dict in Python).
  // Typical examples are threads, timeouts etc. Can be empty.
  map<string, string> parameters = 1;
  // A decomposed specification.
  DecompLTLSpecification decomp_specification = 2;
  // AIGER circuit.
  optional AigerCircuit circuit = 3;
  // Shows whether the specification is realizable or unrealizable.
  bool realizable = 4;
}

message LTLMCSolution {
  // A status that includes useful information about the run.
  LTLMCStatus status = 1;
  // which tool has created the response.
  Tool tool = 2;
  // A trace, proving the violation of the specification.
  optional Trace counterexample = 3;
  // How long the tool took to create the result.
  optional google.protobuf.Duration time = 4;
}    \end{lstlisting}
    \caption{Parts of the protocol buffer interfaces. Some messages and definitions are missing for simplification.}
    \label{app:protobuf}
\end{figure}

\clearpage

\section{Distribution of \texttt{test}, \texttt{large}, and \texttt{syntcomp} dataset over input size}

\begin{figure}[!ht]
  \includegraphics[width=\textwidth]{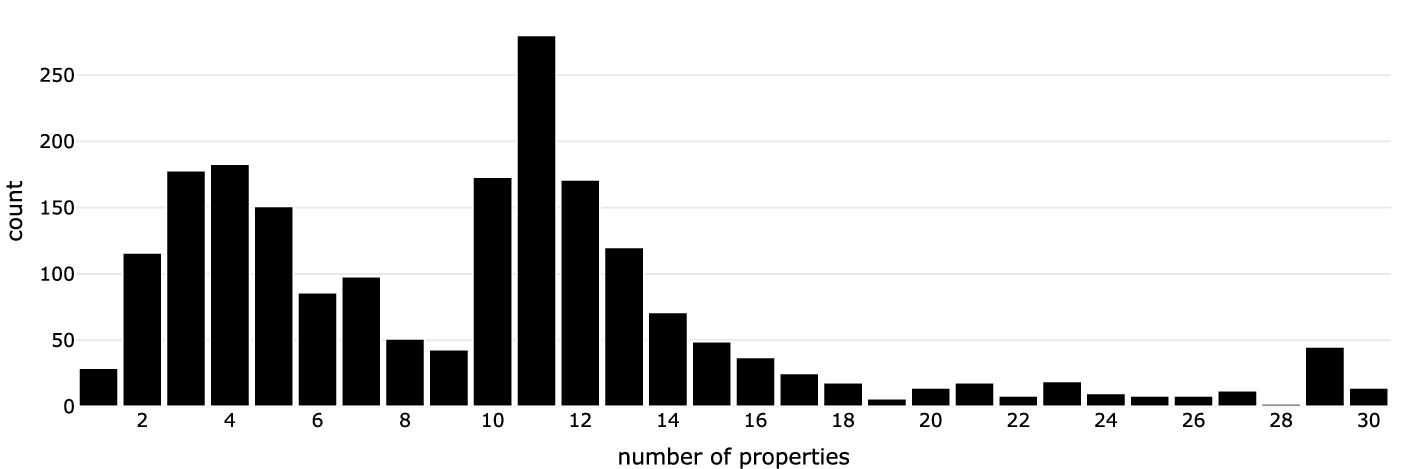}
  \includegraphics[width=\textwidth]{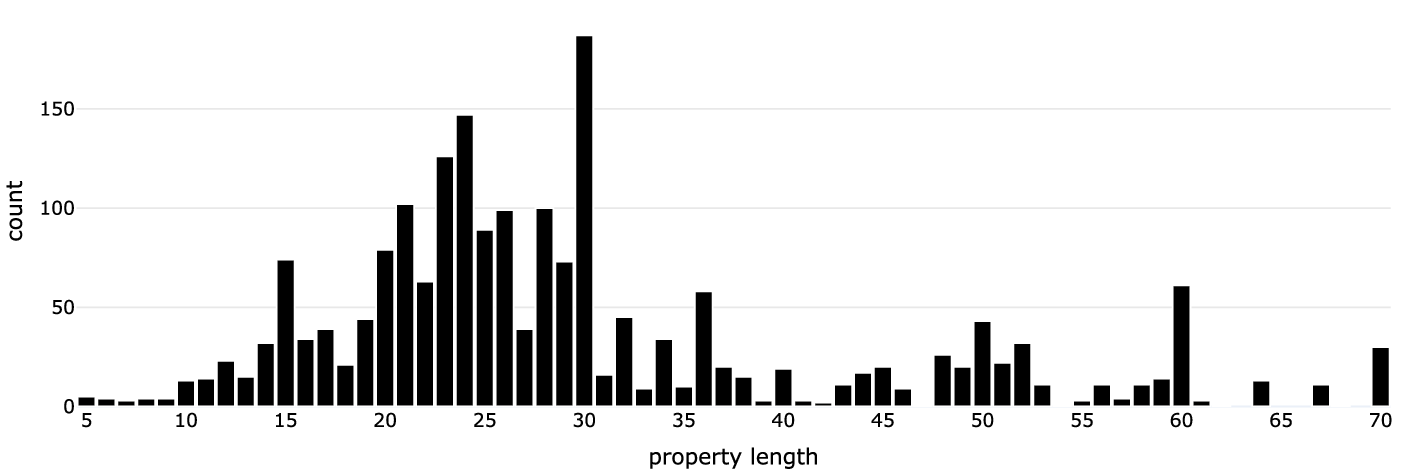}
  \caption{Distribution of the joint dataset of \texttt{large} and \texttt{test} over the number of properties in a sample and the size of the largest property in each sample.}
  \label{app:dist_large}
\end{figure}

\begin{figure}[!ht]
  \centering
  \includegraphics[width=\textwidth]{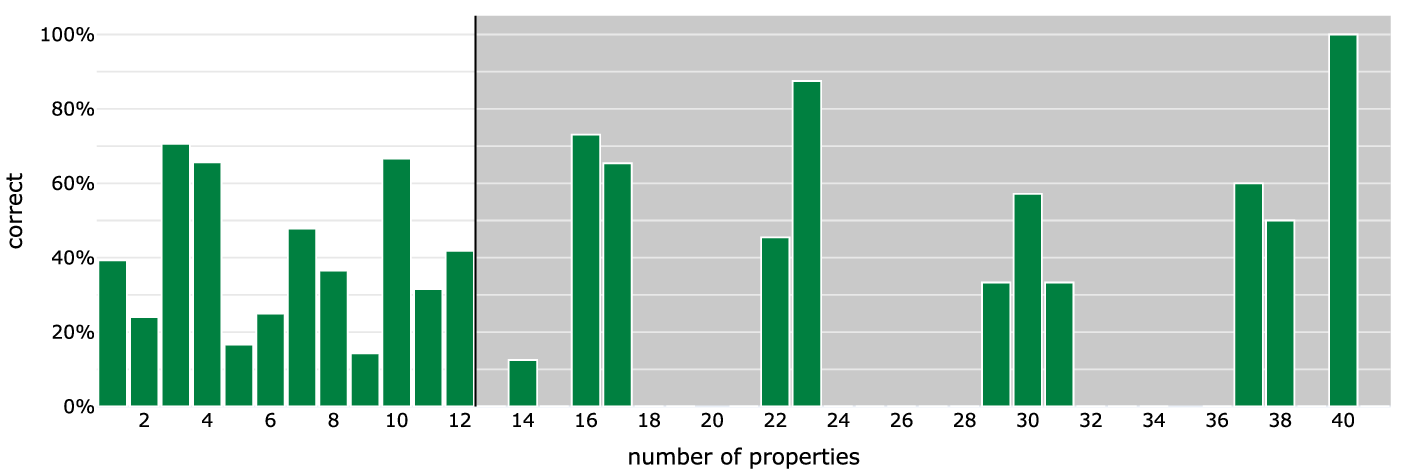}
  \includegraphics[width=\textwidth]{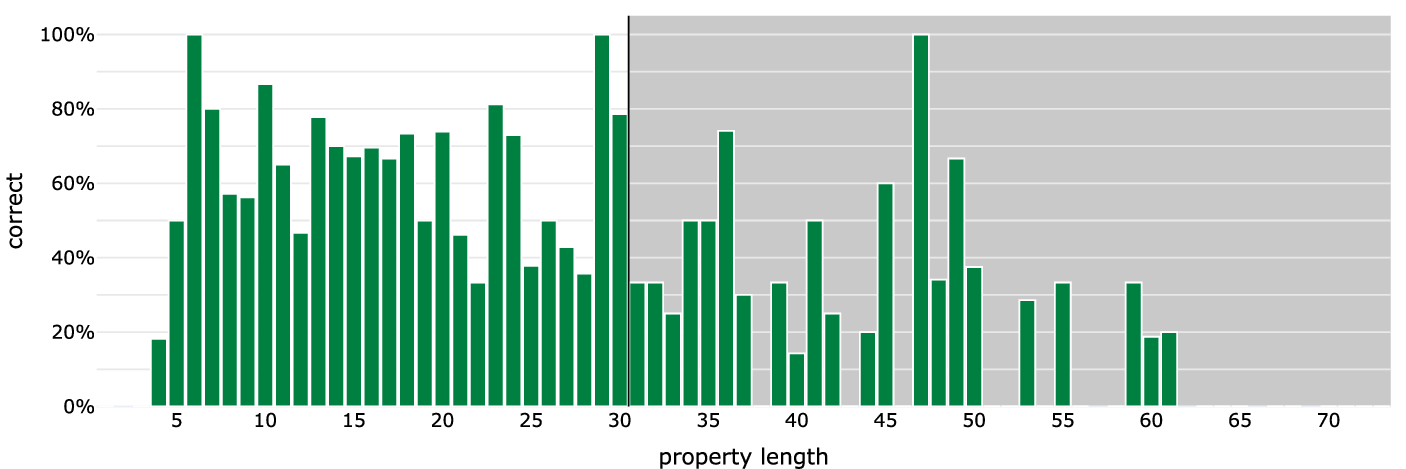}
  \label{app:share_syntcomp}

  \includegraphics[width=\textwidth]{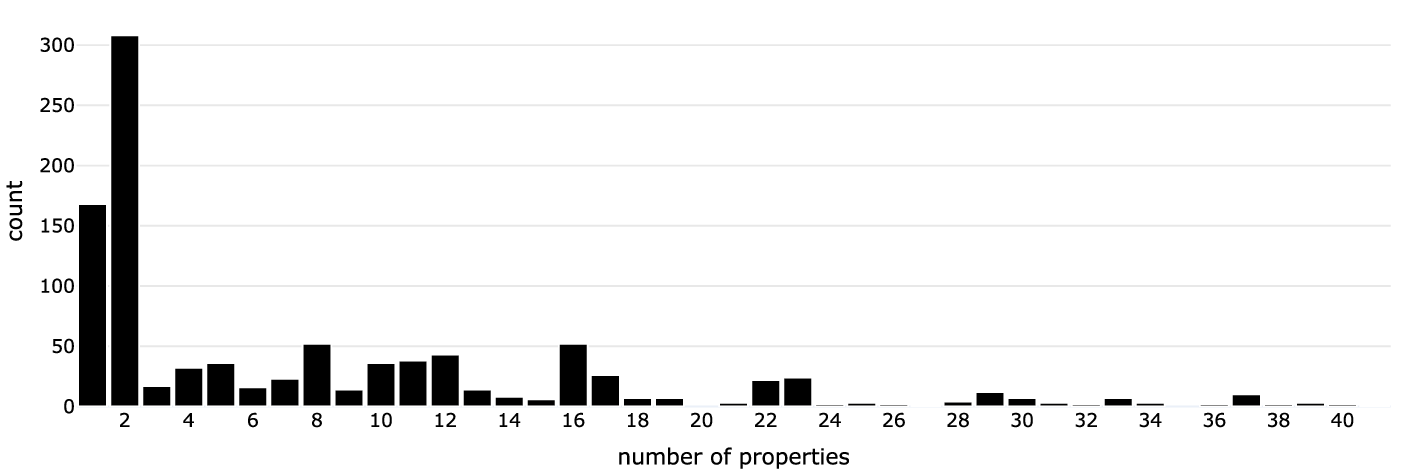}
  \includegraphics[width=\textwidth]{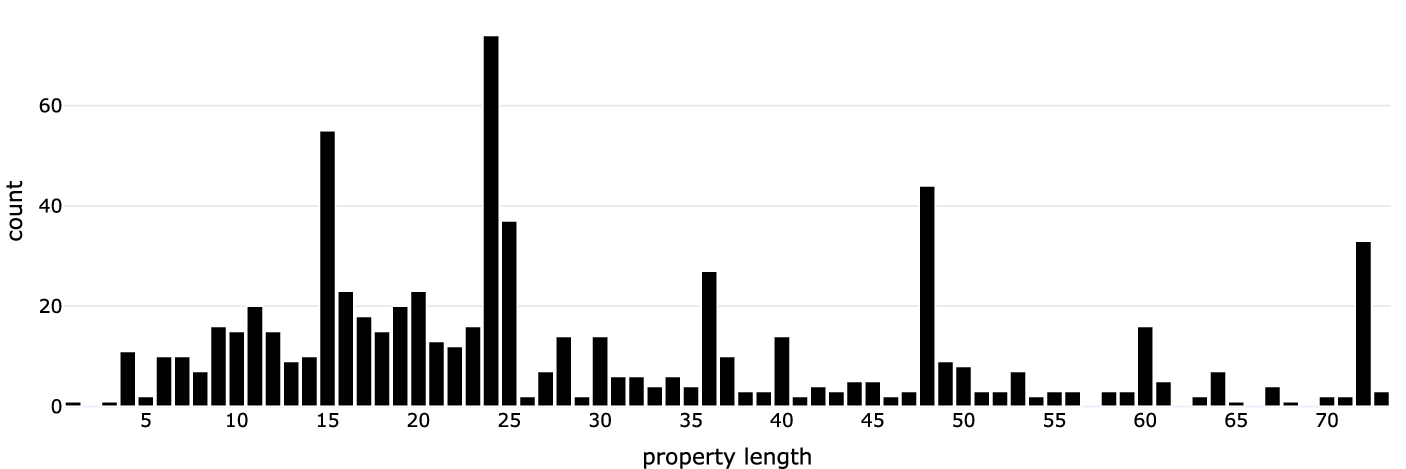}
  \caption{Top: Share of correct solutions on the \texttt{syntcomp} datasets over the number of properties in a sample and the size of the largest property in each sample. A darker background indicates sizes larger than seen during training. Bottom: Distribution of \texttt{syntcomp} dataset over the number of properties in a sample and the size of the largest property in each sample.}
  \label{app:dist_syntcomp}
\end{figure}

\clearpage

\section{Syntcomp Results}

Here, we provide detailed results on the number of solved instances (Table~\ref{app:syntcomp-performance-exclusive}), the average times per tool (Table~\ref{app:times}), and the sizes of the solutions (Table~\ref{app:sizes}).

\begin{table}[!ht]
  \caption{Performance of several tools on the 2022 SYNTCOMP benchmarks. Total number of samples: 1075. Tools denoted with $\dagger$ include experiments on own hardware. Other results are solely based on the SYNTCOMP results. Exclusive solves are instances where no other listed tool provided a solution}
  \setlength{\tabcolsep}{0.6em}
  \label{app:syntcomp-performance-exclusive}
  \begin{tabular}{@{}llll@{}}
    \toprule
    \multirow{2}{*}{tool}   & total solved       & total solved           & exclusively solved     \\
                            & best configuration & grouped configurations & grouped configurations \\ \midrule
    Strix$^\dagger$         & 858                & $919$                  & $116$                  \\
    sdf                     & 719                & $723$                  & $17$                   \\
    ltlsynt                 & 707                & $717$                  & $5$                    \\
    Otus                    & 502                & $503$                  & $0$                    \\
    Neural Solver$^\dagger$ & 374                & $402$                  & $20$                   \\
    BoSy$^\dagger$          & 347                & $347$                  & $0$                    \\ \bottomrule
    \end{tabular}
  \end{table}

  \begin{table}[!ht]
    \caption{Average times of Strix, BoSy, and the neural solver on different hardware configurations. Averaged over all SYNTCOMP instances. For the symbolic solvers, we additionally report the number of solved samples for the different hardware configurations. The Neural Solvers accuracy does not depend on the hardware configuration. Experiments on the Apple Mac Book Pro (MBP: M1 Max, 10 cores, 32 Neural Cores) are 20 randomly selected instances from SYNTCOMP. Given the low standard deviation of the neural solver that we found in other experiments, this is an accurate representation.}
    \setlength{\tabcolsep}{0.6em}
    \label{app:times}
    \centering
    \begin{tabular}{@{}llll@{}}
      \toprule
      Experiment                            & mean (s) & std (s) & solved instances\\ \midrule
      Strix on CPU cluster 120s timeout     & $5.1$    & $14.3$  & 821\\
      Strix on GPU cluster 120s timeout     & $3.7$    & $12.4$  & 833\\
      Strix on CPU cluster $1h$ timeout     & $33.4$   & $185.3$ & 858\\
      Strix-ltl\_synth\_zlk\_bfs (SYNTCOMP) & $28.2$   & $145.7$ & 816\\
      BoSy on CPU cluster 120s timeout      & $19.4 $  & $28.1$  & 347\\
      Neural Solver on GPU cluster          & $8.6$    & $3.3 $  & \\
      \;\;$\rightarrow$ thereof model-checking  & $0.35$   & $0.93$  & \\
      Neural Solver on CPU cluster          & $79.4$   & $28$   &  \\
      \;\;$\rightarrow$ thereof model-checking  & $0.66$   & $1.3$   & \\
      Neural Solver on MBP                  & $17.8$   & -       & \\
      \;\;$\rightarrow$ thereof model-checking  & $0.59$   & -       & \\ \bottomrule
      \end{tabular}
    \end{table}

    \begin{table}[!ht]
      \caption{Comparison of the number of latches of different tools. Each symbolic solver is compared with the neural solver on all instances that both solvers solve correctly. Calculations on $\dagger$ marked tools are solely based on the realizable fraction of SYNTCOMP results, as SYNTCOMP only reports circuits for realizable instances.}
      \setlength{\tabcolsep}{0.6em}
      \label{app:sizes}
      \centering
      \begin{tabular}{@{}llll@{}}
        \toprule
        Symbolic Solver   & \begin{tabular}[c]{@{}l@{}}average \\ Symbolic Solver\end{tabular} & \begin{tabular}[c]{@{}l@{}}average \\ Neural Solver\end{tabular} & \begin{tabular}[c]{@{}l@{}}Neural Solver \\ smaller by (\%)\end{tabular} \\ \midrule
        Strix             & 1.4                                                                & 0.64                                                             & 54.9                                                                     \\
        BoSy              & 1.11                                                               & 0.64                                                             & 42.3                                                                     \\
        ltlsynt$^\dagger$ & 2.49                                                               & 1.05                                                             & 57.7                                                                     \\
        Otus$^\dagger$    & 13.21                                                              & 1.05                                                             & 92                                                                       \\
        sdf$^\dagger$     & 44.84                                                              & 1.05                                                             & 97.6                                                                     \\ \bottomrule
        \end{tabular}
      \end{table}

\section{Example Instances}
\label{app:samples}

Figure~\ref{fig:parameterized_arbiter} shows an example instance in which the neural solver found a solution, but Strix did not within the time limit.
Figure~\ref{fig:val_12} shows an example instance on which the Transformer neural network was trained on, consisting of $12$ properties.
Figure~\ref{fig:large_30} shows an example instance of the \texttt{large} data set, which was solved by the neural solver.
Figure~\ref{fig:syntcomp_40} shows an example instance from SYNTCOMP, which was solved by the neural solver (and Strix).

\begin{figure}
  \centering
  \scriptsize
  \begin{lstlisting}[language=json]
{
"inputs": ["i0", "i1", "i2", "i3", "i4"],
"outputs": ["o0", "o1", "o2", "o3", "o4"],
"assumptions": [
  "(G ((! (o0)) | (X (((! (i0)) & (! (i2))) U ((! (i0)) & (i2))))))",
  "(G ((! (o2)) | (X (((! (i1)) & (! (i0))) U ((! (i1)) & (i0))))))"
],
"guarantees": [
  "(G (F (o3)))",
  "(G ((o0) -> (X ((i0) R (((i0) -> (o2)) & ((! (i0)) -> (o4)))))))",
  "(G (((i3) & (X (i0))) -> (F ((o2) & (o0)))))",
  "(G ((i0) -> (F (o3))))",
  "(G (((i2) & (X (i3))) -> (X (X (X (X (X (X (X (X (X (X ((o3) & (o0))))))))))))))",
  "(G ((o1) -> (X (((! (o1)) U (i3)) | (G (! (o1)))))))",
  "(G (((i3) & (X (i0))) -> ((X (o1)) <-> (X (i2)))))",
  "(G ((o3) -> (X (((! (o3)) U (i1)) | (G (! (o3)))))))",
  "(G (((i4) & (o1)) -> (X (X (X (X (X (o4))))))))",
  "(G ((X (o4)) -> (i0)))"
]
}
  \end{lstlisting}
  \caption{Example instance from the training set with 12 properties.}
  \label{fig:val_12}
\end{figure}

\begin{figure}
    \centering
    \scriptsize
    \begin{lstlisting}[language=json]
INFO {
  TITLE:       "Full Arbiter, unrealizable variant 1"
  DESCRIPTION: "Parameterized Arbiter, where no spurious grants are allowed"
  SEMANTICS:   Moore
  TARGET:      Mealy
}
{
  "inputs": ["r_0","r_1","r_2"],
  "outputs": ["g_0","g_1","g_2"],
  "assumptions": [],
  "guarantees": [
    "(G (((g_0) & (G (! (r_0)))) -> (F (! (g_0)))))",
    "(G (((g_0) & (X ((! (r_0)) & (! (g_0))))) -> (X ((r_0) R (! (g_0))))))",
    "(G (((g_1) & (G (! (r_1)))) -> (F (! (g_1)))))",
    "(G (((g_1) & (X ((! (r_1)) & (! (g_1))))) -> (X ((r_1) R (! (g_1))))))",
    "(G (((g_2) & (G (! (r_2)))) -> (F (! (g_2)))))",
    "(G (((g_2) & (X ((! (r_2)) & (! (g_2))))) -> (X ((r_2) R (! (g_2))))))",
    "(G (((! (g_0)) & (! (g_1))) | (((! (g_0)) | (! (g_1))) & (! (g_2)))))",
    "(G (((r_0) & (X (r_1))) -> (X (X (X (X (X (X (X (X (X (X (X (X (X (X (X ((g_0) & (g_1)))))))))))))))))))",
    "(G (((r_0) & (X (r_2))) -> (X (X (X (X (X (X (X (X (X (X (X (X (X (X (X ((g_0) & (g_2)))))))))))))))))))",
    "(G (((r_1) & (X (r_2))) -> (X (X (X (X (X (X (X (X (X (X (X (X (X (X (X ((g_1) & (g_2)))))))))))))))))))",
    "((r_0) R (! (g_0)))",
    "(G ((r_0) -> (F (g_0))))",
    "((r_1) R (! (g_1)))",
    "(G ((r_1) -> (F (g_1))))",
    "((r_2) R (! (g_2)))",
    "(G ((r_2) -> (F (g_2))))"
  ]
}
\end{lstlisting}
    \caption{An instance of a parameterized arbiter without spurious grant that is solved by \tool, but not by any of the symbolic solvers.}
    \label{fig:parameterized_arbiter}
\end{figure}

\begin{figure}
    \centering
    \scriptsize
    \begin{lstlisting}[language=json]
{
  "inputs": ["i0","i1","i10","i11","i12","i13","i14","i2","i3","i4","i6","i7","i8","i9"],
  "outputs": ["o0","o1","o10","o11","o12","o13","o2","o3","o4","o5","o6","o7","o8","o9"],
  "assumptions": [
    "G ( F ( i12 ) )",
    "X ( G ( ( ! ( o1 ) ) | ( ( ( ! ( i13 ) ) & ( ! ( i10 ) ) ) U ( ( ! ( i13 ) ) & ( i10 ) ) ) ) )",
    "X ( G ( ( ! ( o13 ) ) | ( ( ( ! ( i10 ) ) & ( ! ( i11 ) ) ) U ( ( ! ( i10 ) ) & ( i11 ) ) ) ) )"
  ],
  "guarantees": [
    "G ( ( ( i6 ) & ( X ( i13 ) ) ) -> ( X ( X ( X ( X ( X ( X ( X ( X ( X ( ( o1 ) & ( o2 ) ) ) ) ) ) ) ) ) ) ) )",
    "G ( ( i13 ) -> ( F ( o7 ) ) )",
    "G ( ( o13 ) -> ( X ( ( o13 ) | ( o3 ) ) ) )",
    "G ( ( i0 ) -> ( F ( o1 ) ) )",
    "G ( ( i7 ) -> ( ( o2 ) <-> ( X ( o0 ) ) ) )",
    "o3",
    "G ( ( ( o0 ) | ( ( o13 ) & ( i11 ) ) ) -> ( X ( F ( ( ( o12 ) | ( o13 ) ) R ( i12 ) ) ) ) )",
    "G ( ( ( ( ( ( ( ( ( ! ( i10 ) ) & ( i2 ) ) & ( ! ( i13 ) ) ) & ( ! ( i8 ) ) ) & ( ! ( i11 ) ) ) & ( ! ( i9 ) ) ) & ( ! ( i6 ) ) ) & ( i12 ) ) -> ( ( o2 ) <-> ( i0 ) ) )",
    "G ( ( ( i2 ) & ( X ( i4 ) ) ) -> ( X ( X ( X ( X ( X ( X ( X ( ( o1 ) & ( o2 ) ) ) ) ) ) ) ) ) )",
    "( i12 ) R ( ! ( o9 ) )",
    "G ( ( ( ( ( ( ( ! ( i6 ) ) & ( i13 ) ) & ( ! ( i0 ) ) ) & ( i11 ) ) & ( ! ( i4 ) ) ) & ( i10 ) ) -> ( F ( ( ( ( ( ( ( ! ( o3 ) ) & ( ! ( o6 ) ) ) & ( o9 ) ) & ( ! ( o0 ) ) ) & ( o2 ) ) & ( ! ( o7 ) ) ) & ( o13 ) ) ) )",
    "G ( ( i11 ) -> ( ( X ( ( ( ! ( o4 ) ) & ( o9 ) ) & ( ! ( o6 ) ) ) ) <-> ( i12 ) ) )",
    "G ( ( ( ( ( ( ( o0 ) & ( ! ( o8 ) ) ) <-> ( ( ! ( o8 ) ) | ( o0 ) ) ) & ( ! ( o4 ) ) ) & ( ! ( o9 ) ) ) <-> ( ( ( ( ( o4 ) & ( ! ( o9 ) ) ) <-> ( ( o9 ) & ( ! ( o4 ) ) ) ) | ( o0 ) ) | ( o8 ) ) ) & ( ( ( ( ( ( o3 ) & ( ! ( o6 ) ) ) <-> ( ( ! ( o6 ) ) | ( o3 ) ) ) & ( ! ( o12 ) ) ) & ( ! ( o2 ) ) ) <-> ( ( ( ( ( o12 ) & ( ! ( o2 ) ) ) <-> ( ( o2 ) & ( ! ( o12 ) ) ) ) | ( o3 ) ) | ( o6 ) ) ) )",
    "G ( ( ! ( o12 ) ) | ( ! ( o2 ) ) )",
    "G ( ( ( ( ( ( ! ( o3 ) ) & ( o5 ) ) & ( o12 ) ) & ( ! ( o4 ) ) ) & ( G ( ( ( ( i13 ) | ( ! ( i14 ) ) ) | ( ! ( i0 ) ) ) | ( i9 ) ) ) ) -> ( F ( ( ( ( o3 ) | ( ! ( o5 ) ) ) | ( ! ( o12 ) ) ) | ( o4 ) ) ) )",
    "o10",
    "F ( ( i8 ) & ( ( ! ( i7 ) ) U ( ( i4 ) & ( ! ( i7 ) ) ) ) )",
    "G ( ( i12 ) -> ( F ( o0 ) ) )",
    "( ( ! ( i11 ) ) & ( G ( ( ( ( i11 ) & ( ! ( o8 ) ) ) -> ( X ( i11 ) ) ) & ( ( o8 ) -> ( X ( ! ( i11 ) ) ) ) ) ) ) -> ( ( ! ( o12 ) ) & ( G ( ( ( ( ( ! ( i11 ) ) & ( X ( i11 ) ) ) -> ( X ( ( ! ( o8 ) ) & ( X ( F ( o8 ) ) ) ) ) ) & ( ( ( ! ( o8 ) ) & ( X ( ! ( i11 ) ) ) ) -> ( X ( ! ( o8 ) ) ) ) ) & ( ( ! ( o12 ) ) | ( ! ( o8 ) ) ) ) ) )",
    "G ( ( o0 ) -> ( X ( ( ( ! ( o0 ) ) U ( i3 ) ) | ( G ( ! ( o0 ) ) ) ) ) )",
    "G ( ( ( i9 ) & ( X ( i8 ) ) ) -> ( X ( X ( X ( X ( X ( X ( X ( X ( ( o5 ) & ( o2 ) ) ) ) ) ) ) ) ) ) )",
    "G ( F ( ( ! ( i13 ) ) | ( X ( o12 ) ) ) )",
    "G ( ( ( ( ( ( ( ( ! ( i12 ) ) & ( i1 ) ) & ( i10 ) ) & ( ! ( i11 ) ) ) & ( i6 ) ) & ( ! ( i9 ) ) ) & ( ! ( i0 ) ) ) -> ( ( o8 ) <-> ( i3 ) ) )",
    "G ( ( ( i13 ) & ( X ( i6 ) ) ) -> ( X ( X ( X ( X ( X ( X ( ( o11 ) & ( o8 ) ) ) ) ) ) ) ) )",
    "G ( F ( ( ! ( i0 ) ) | ( X ( o1 ) ) ) )",
    "G ( ( ! ( o4 ) ) | ( ! ( o9 ) ) )",
    "G ( ( ( ! ( o7 ) ) & ( ! ( o2 ) ) ) | ( ( ( ! ( o7 ) ) | ( ! ( o2 ) ) ) & ( ! ( o5 ) ) ) )"
  ]
}
    \end{lstlisting}
    \caption{Solved example instance with $30$ properties taken from \texttt{large}.}
    \label{fig:large_30}
\end{figure}

\begin{figure}
    \centering
    \scriptsize
    \begin{lstlisting}[language=json]
INFO {
  TITLE:       "Full Arbiter, unrealizable variant 2"
  DESCRIPTION: "Parameterized Arbiter, where no spurious grants are allowed"
  SEMANTICS:   Mealy
  TARGET:      Mealy
}

{
  "inputs": ["r_0","r_1","r_2","r_3","r_4","r_5"],
  "outputs": ["g_0","g_1","g_2","g_3","g_4","g_5"],
  "assumptions": [],
  "guarantees": [
    "(G (((g_0) & (G (! (r_0)))) -> (F (! (g_0)))))",
    "(G (((g_0) & (X ((! (r_0)) & (! (g_0))))) -> (X ((r_0) R (! (g_0))))))",
    "(G (((g_1) & (G (! (r_1)))) -> (F (! (g_1)))))",
    "(G (((g_1) & (X ((! (r_1)) & (! (g_1))))) -> (X ((r_1) R (! (g_1))))))",
    "(G (((g_2) & (G (! (r_2)))) -> (F (! (g_2)))))",
    "(G (((g_2) & (X ((! (r_2)) & (! (g_2))))) -> (X ((r_2) R (! (g_2))))))",
    "(G (((g_3) & (G (! (r_3)))) -> (F (! (g_3)))))",
    "(G (((g_3) & (X ((! (r_3)) & (! (g_3))))) -> (X ((r_3) R (! (g_3))))))",
    "(G (((g_4) & (G (! (r_4)))) -> (F (! (g_4)))))",
    "(G (((g_4) & (X ((! (r_4)) & (! (g_4))))) -> (X ((r_4) R (! (g_4))))))",
    "(G (((g_5) & (G (! (r_5)))) -> (F (! (g_5)))))",
    "(G (((g_5) & (X ((! (r_5)) & (! (g_5))))) -> (X ((r_5) R (! (g_5))))))",
    "(G (((((! (g_0)) & (! (g_1))) & (! (g_2))) & (((! (g_3)) & (! (g_4))) | (((! (g_3)) | (! (g_4))) & (! (g_5))))) | ((((((! (g_0)) & (! (g_1))) | (((! (g_0)) | (! (g_1))) & (! (g_2)))) & (! (g_3))) & (! (g_4))) & (! (g_5)))))",
    "(G (((r_0) & (X (r_1))) -> (F ((g_0) & (g_1)))))",
    "(G (((r_0) & (X (r_2))) -> (F ((g_0) & (g_2)))))",
    "(G (((r_0) & (X (r_3))) -> (F ((g_0) & (g_3)))))",
    "(G (((r_0) & (X (r_4))) -> (F ((g_0) & (g_4)))))",
    "(G (((r_0) & (X (r_5))) -> (F ((g_0) & (g_5)))))",
    "(G (((r_1) & (X (r_2))) -> (F ((g_1) & (g_2)))))",
    "(G (((r_1) & (X (r_3))) -> (F ((g_1) & (g_3)))))",
    "(G (((r_1) & (X (r_4))) -> (F ((g_1) & (g_4)))))",
    "(G (((r_1) & (X (r_5))) -> (F ((g_1) & (g_5)))))",
    "(G (((r_2) & (X (r_3))) -> (F ((g_2) & (g_3)))))",
    "(G (((r_2) & (X (r_4))) -> (F ((g_2) & (g_4)))))",
    "(G (((r_2) & (X (r_5))) -> (F ((g_2) & (g_5)))))",
    "(G (((r_3) & (X (r_4))) -> (F ((g_3) & (g_4)))))",
    "(G (((r_3) & (X (r_5))) -> (F ((g_3) & (g_5)))))",
    "(G (((r_4) & (X (r_5))) -> (F ((g_4) & (g_5)))))",
    "((r_0) R (! (g_0)))",
    "(G ((r_0) -> (F (g_0))))",
    "((r_1) R (! (g_1)))",
    "(G ((r_1) -> (F (g_1))))",
    "((r_2) R (! (g_2)))",
    "(G ((r_2) -> (F (g_2))))",
    "((r_3) R (! (g_3)))",
    "(G ((r_3) -> (F (g_3))))",
    "((r_4) R (! (g_4)))",
    "(G ((r_4) -> (F (g_4))))",
    "((r_5) R (! (g_5)))",
    "(G ((r_5) -> (F (g_5))))"
  ]
}
    \end{lstlisting}
    \caption{Solved example instance from SYNTCOMP with $40$ properties.}
    \label{fig:syntcomp_40}
\end{figure}

\end{document}